\documentclass[hyphens,conference]{IEEEtran}

\newenvironment{acks}{\section*{Acknowledgments}}{}

\newcommand{\swallow}[1]{}

\newcommand{\colorcomment}[2]{\relax}

\usepackage{paralist}
\usepackage[utf8]{inputenc}
\usepackage{amsmath}
\usepackage{booktabs}
\usepackage{comment}
\usepackage{siunitx}
\usepackage{multirow}
\usepackage{array}
\usepackage{makecell}
\let\labelindent\relax
\usepackage[inline]{enumitem}
\usepackage{pgfplotstable}
\usepackage{xspace}
\usepackage{color}
\let\chapter\section 

\usepackage{hyperref}
\definecolor{linkcolor}{rgb}{0.65,0,0}
\definecolor{citecolor}{rgb}{0,0.4,0}
\definecolor{urlcolor}{rgb}{0,0,0.65}
\hypersetup{hyperindex=true,pdfpagemode=UseNone,pdfstartview=FitH,colorlinks=true, linkcolor=linkcolor, urlcolor=urlcolor, citecolor=citecolor}
\usepackage[nameinlink,capitalise,noabbrev]{cleveref}
\usepackage[numbers,sort]{natbib}
\usepackage{stfloats}
\usepackage{balance}
\usepackage[ruled,vlined,linesnumbered]{algorithm2e}

\usepackage{tikz}
\usetikzlibrary{fit,calc}
\newcommand*{\tikzmk}[1]{\tikz[remember picture,overlay,] \node (#1) {};\ignorespaces}
\newcommand{\boxit}[1]{\tikz[remember picture,overlay]{\node[yshift=1.5pt,fill=#1,opacity=.25,fit={(A)($(B)+(.75\linewidth,.8\baselineskip)$)}] {};}\ignorespaces}

\usepackage{tikzscale}
\usepackage{pgfplots}
\usepackage{pgfplotstable}
\usepgflibrary{fpu}
\usetikzlibrary{positioning}
\usetikzlibrary{shapes}
\usetikzlibrary{arrows}
\usetikzlibrary{arrows.meta}
\usetikzlibrary{fit}
\usetikzlibrary{calc}
\usetikzlibrary{patterns}
\definecolor{color1}{HTML}{FFA978} 
\definecolor{color2}{HTML}{C84526} 
\definecolor{color3}{HTML}{340701}

\crefname{lstlisting}{listing}{listings}
\Crefname{lstlisting}{Listing}{Listings}
\DeclareSIUnit{\fahrenheit}{\degree F}
\newcolumntype{M}[1]{>{\centering\arraybackslash}m{#1}} 
\newcolumntype{R}[1]{>{\raggedleft\arraybackslash }b{#1}}
\newcolumntype{L}[1]{>{\raggedright\arraybackslash }b{#1}}
\newcolumntype{C}[1]{>{\centering\arraybackslash }b{#1}}
\newcommand{\parhead}[1]{\vspace{3pt plus 1pt minus 1pt}\par\noindent\textbf{#1}\hspace{.75em minus .5em}}
\newcommand{\etal}{et~al.\xspace} 
 
\newcommand{\eg}{e.g.,\ }

\newcommand{\WebGLtwo}{WebGL\,2.0\xspace}
\newcommand{\GenIII}{\textsc{Gen\,3}\xspace}
\newcommand{\GenIV}{\textsc{Gen\,4}\xspace}
\newcommand{\GenVIII}{\textsc{Gen\,8}\xspace}
\newcommand{\GenX}{\textsc{Gen\,10}\xspace}

\newcommand{\performancenow}{\texttt{performance.now()}\xspace}

\newcommand{\AmIUnique}{\textsc{AmIUnique}\xspace}

\newcommand{\baserate}{\mathit{base rate}}

\newcommand{\oneMP}{\textsf{1MP}}
\newcommand{\twoMP}{\textsf{2MP}}
\newcommand{\threeMP}{\textsf{3MP}}
\newcommand{\rest}{\mathit{rest}}

\usepackage{listings}
\definecolor{lightgray}{rgb}{.95,.95,.95}
\definecolor{darkgray}{rgb}{.4,.4,.4}
\definecolor{purple}{rgb}{0.65, 0.12, 0.82}
\definecolor{dkgreen}{rgb}{0,0.6,0}
\definecolor{lightgreen}{rgb}{0.2,0.8,0.4}
\definecolor{lightcyan}{rgb}{0,0.8,0.8}

\lstdefinelanguage{JavaScript}{
  keywords={typeof, new, true, false, catch, function, return, null, catch, switch, var, if, in, while, do, else, case, break},
  keywordstyle=\color[rgb]{0,0,0.75},
  ndkeywords={class, export, boolean, throw, implements, import, this},
  ndkeywordstyle=\color{darkgray}\bfseries,
  identifierstyle=\color{black},
  sensitive=false,
  comment=[l]{//},
  morecomment=[s]{/*}{*/},
  commentstyle=\color{purple}\ttfamily,
  stringstyle=\color{red}\ttfamily,
  morestring=[b]',
  morestring=[b]"
}

\lstdefinelanguage{HTML5}{
  language=html,
  sensitive=true,	
  alsoletter={<>=-},	
  morecomment=[s]{<!-}{-->},
  tag=[s],
  otherkeywords={
	<!DOCTYPE,
  </html, <html, <head, <title, </title, <style, </style, <link, </head, <meta, />,
	</body, <body,
	</div, <div, </div>, 
	</p, <p, </p>,
	</script>, <script>,
  <canvas, /canvas>, <svg, <rect, <animateTransform, </rect>, </svg>, <video, <source, <iframe, ></iframe>, </video>, <image, </image>, <header, </header, <article, </article
  },
  ndkeywords={
  =,
  charset=, src=, id=, width=, height=, style=, type=, rel=, href=,
  fill=, attributeName=, begin=, dur=, from=, to=, poster=, controls=, x=, y=, repeatCount=, xlink:href=,
  margin:, padding:, background-image:, border:, top:, left:, position:, width:, height:, margin-top:, margin-bottom:, font-size:, line-height:,
  transform:, -moz-transform:, -webkit-transform:,
  animation:, -webkit-animation:,
  transition:,  transition-duration:, transition-property:, transition-timing-function:,
  }
}

\lstdefinestyle{web} {
  basicstyle={\footnotesize\ttfamily},   
  frame=single,
  xleftmargin={0.75cm},
  numbers=left,
  stepnumber=1,
  firstnumber=1,
  numberfirstline=true,	
  identifierstyle=\color{black},
  keywordstyle=\color{blue}\bfseries,
  ndkeywordstyle=\color{editorGreen}\bfseries,
  stringstyle=\color{editorOcher}\ttfamily,
  commentstyle=\color{purple}\ttfamily,
  language=HTML5,
  alsolanguage=JavaScript,
  alsodigit={.:;},	
  tabsize=2,
  showtabs=false,
  showspaces=false,
  showstringspaces=false,
  extendedchars=true,
  breaklines=true,
  literate=%
  {Ö}{{\"O}}1
  {Ä}{{\"A}}1
  {Ü}{{\"U}}1
  {ß}{{\ss}}1
  {ü}{{\"u}}1
  {ä}{{\"a}}1
  {ö}{{\"o}}1
}

\lstdefinelanguage{GLSL}
{
	sensitive=true,
	alsoletter={\#},
	morekeywords=[1]{
		attribute, const, uniform, varying,
		layout, centroid, flat, smooth,
		noperspective, break, continue, do,
		for, while, switch, case, default, if,
		else, in, out, inout, float, int, void,
		bool, true, false, invariant, discard,
		return, mat2, mat3, mat4, mat2x2, mat2x3,
		mat2x4, mat3x2, mat3x3, mat3x4, mat4x2,
		mat4x3, mat4x4, vec2, vec3, vec4, ivec2,
		ivec3, ivec4, bvec2, bvec3, bvec4, uint,
		uvec2, uvec3, uvec4, lowp, mediump, highp,
		precision, sampler1D, sampler2D, sampler3D,
		samplerCube, sampler1DShadow,
		sampler2DShadow, samplerCubeShadow,
		sampler1DArray, sampler2DArray,
		sampler1DArrayShadow, sampler2DArrayShadow,
		isampler1D, isampler2D, isampler3D,
		isamplerCube, isampler1DArray,
		isampler2DArray, usampler1D, usampler2D,
		usampler3D, usamplerCube, usampler1DArray,
		usampler2DArray, sampler2DRect,
		sampler2DRectShadow, isampler2DRect,
		usampler2DRect, samplerBuffer,
		isamplerBuffer, usamplerBuffer, sampler2DMS,
		isampler2DMS, usampler2DMS,
		sampler2DMSArray, isampler2DMSArray,
		usampler2DMSArray, struct
	},
	morekeywords=[2]{
		radians,degrees,sin,cos,tan,asin,acos,atan,
		atan,sinh,cosh,tanh,asinh,acosh,atanh,pow,
		exp,log,exp2,log2,sqrt,inversesqrt,abs,sign,
		floor,trunc,round,roundEven,ceil,fract,mod,modf,
		min,max,clamp,mix,step,smoothstep,isnan,isinf,
		floatBitsToInt,floatBitsToUint,intBitsToFloat,
		uintBitsToFloat,length,distance,dot,cross,
		normalize,faceforward,reflect,refract,
		matrixCompMult,outerProduct,transpose,
		determinant,inverse,lessThan,lessThanEqual,
		greaterThan,greaterThanEqual,equal,notEqual,
		any,all,not,textureSize,texture,textureProj,
		textureLod,textureOffset,texelFetch,
		texelFetchOffset,textureProjOffset,
		textureLodOffset,textureProjLod,
		textureProjLodOffset,textureGrad,
		textureGradOffset,textureProjGrad,
		textureProjGradOffset,texture1D,texture1DProj,
		texture1DProjLod,texture2D,texture2DProj,
		texture2DLod,texture2DProjLod,texture3D,
		texture3DProj,texture3DLod,texture3DProjLod,
		textureCube,textureCubeLod,shadow1D,shadow2D,
		shadow1DProj,shadow2DProj,shadow1DLod,
		shadow2DLod,shadow1DProjLod,shadow2DProjLod,
		dFdx,dFdy,fwidth,noise1,noise2,noise3,noise4,
		EmitVertex,EndPrimitive
	},
	morekeywords=[3]{
		gl_VertexID,gl_InstanceID,gl_Position,
		gl_PointSize,gl_ClipDistance,gl_PerVertex,
		gl_Layer,gl_ClipVertex,gl_FragCoord,
		gl_FrontFacing,gl_ClipDistance,gl_FragColor,
		gl_FragData,gl_MaxDrawBuffers,gl_FragDepth,
		gl_PointCoord,gl_PrimitiveID,
		gl_MaxVertexAttribs,gl_MaxVertexUniformComponents,
		gl_MaxVaryingFloats,gl_MaxVaryingComponents,
		gl_MaxVertexOutputComponents,
		gl_MaxGeometryInputComponents,
		gl_MaxGeometryOutputComponents,
		gl_MaxFragmentInputComponents,
		gl_MaxVertexTextureImageUnits,
		gl_MaxCombinedTextureImageUnits,
		gl_MaxTextureImageUnits,
		gl_MaxFragmentUniformComponents,
		gl_MaxDrawBuffers,gl_MaxClipDistances,
		gl_MaxGeometryTextureImageUnits,
		gl_MaxGeometryOutputVertices,
		gl_MaxGeometryOutputVertices,
		gl_MaxGeometryTotalOutputComponents,
		gl_MaxGeometryUniformComponents,
		gl_MaxGeometryVaryingComponents,gl_DepthRange,
		\#version,core
	},
	morecomment=[l]{//},
	morecomment=[s]{/*}{*/},
    commentstyle=\color{purple}\ttfamily,
	keywordstyle=[1]\color[rgb]{0,0,0.75},
	keywordstyle=[2]\color[rgb]{0.5,0.0,0.0},
	keywordstyle=[3]\color[rgb]{0.127,0.427,0.514},
	stringstyle=\color[rgb]{0.639,0.082,0.082},
}

\AtBeginDocument{%
  \providecommand\BibTeX{{%
    \normalfont B\kern-0.5em{\scshape i\kern-0.25em b}\kern-0.8em\TeX}}}

\newcommand{\DrawnApart}{\textsc{Drawn\-Apart}\xspace}
\newcommand{\FPStalker}{\textsc{FP-Stalker}\xspace}
\newcommand{\pmtxt}{$\pm$}
\newcommand{\sota}{state-of-the-art}

\def\HiLi{\leavevmode\rlap{\hbox to \hsize{\color{yellow!50}\leaders\hrule height .8\baselineskip depth .5ex\hfill}}}
\usepackage{balance}

\begin{document}

\title{\DrawnApart: A Device~Identification~Technique based on Remote~GPU~Fingerprinting} 

\author{
    \IEEEauthorblockN{Tomer Laor*}
    \IEEEauthorblockA{Ben-Gurion Univ.\ of the Negev\\
      tomerlao@post.bgu.ac.il}
	\and
    \IEEEauthorblockN{Naif Mehanna*}
	\IEEEauthorblockA{Univ.~Lille, CNRS, Inria\\
		naif.mehanna@univ-lille.fr}
	\and
	\IEEEauthorblockN{Antonin Durey}
	\IEEEauthorblockA{Univ.~Lille, CNRS, Inria\\
		antonin.durey@univ-lille.fr}
	\and
  \IEEEauthorblockN{Vitaly Dyadyuk}
	\IEEEauthorblockA{Ben-Gurion Univ.\ of the Negev\\
		vitalyd@post.bgu.ac.il}
	\and
	\IEEEauthorblockN{Pierre Laperdrix}
	\IEEEauthorblockA{Univ.~Lille, CNRS, Inria\\
		pierre.laperdrix@univ-lille.fr}
	\and
	\IEEEauthorblockN{Clémentine Maurice}
	\IEEEauthorblockA{Univ.~Lille, CNRS, Inria\\ 
		clementine.maurice@inria.fr}
	\and
	\IEEEauthorblockN{Yossi Oren}
	\IEEEauthorblockA{Ben-Gurion Univ.\ of the Negev\\
		yos@bgu.ac.il}
	\and
	\IEEEauthorblockN{Romain Rouvoy}
	\IEEEauthorblockA{Univ.~Lille, CNRS, Inria / IUF\\
		romain.rouvoy@univ-lille.fr}
	\and
	\IEEEauthorblockN{\hspace{1in}}
	\and
	\IEEEauthorblockN{Walter Rudametkin}
	\IEEEauthorblockA{Univ.~Lille, CNRS, Inria\\
		walter.rudametkin@univ-lille.fr}
	\and
	\IEEEauthorblockN{Yuval Yarom}
	\IEEEauthorblockA{Univ.\ of Adelaide\\
		yval@cs.adelaide.edu.au}
	\and
	\IEEEauthorblockN{\hspace{1in}}
}

\IEEEoverridecommandlockouts
\makeatletter\def\@IEEEpubidpullup{6.5\baselineskip}\makeatother
\IEEEpubid{\parbox{\columnwidth}{
    Network and Distributed Systems Security (NDSS) Symposium 2022\\
    27 February - 3 March 2022, San Diego, CA, USA\\
    ISBN 1-891562-74-6\\
    https://dx.doi.org/10.14722/ndss.2022.24093\\
    www.ndss-symposium.org
}
\hspace{\columnsep}\makebox[\columnwidth]{}}

\maketitle

\begingroup\renewcommand\thefootnote{*}
\footnotetext{Both authors are considered co-first authors.}
\endgroup

\pagestyle{plain}

\begin{abstract}
  Browser fingerprinting aims to identify users or their devices, through scripts that execute in the users' browser and collect information on software or hardware characteristics.
  It is used to track users or as an additional means of identification to improve security.
  Fingerprinting techniques have one significant limitation:
  they are unable to track individual users for an extended duration. 
  This happens because browser fingerprints evolve over time, and these evolutions ultimately cause a fingerprint to be confused with those from other devices sharing similar hardware and software.

  In this paper, we report on a new technique that can significantly extend the tracking time of fingerprint-based tracking methods. 
  Our technique, which we call \DrawnApart, is a new \emph{GPU fingerprinting} technique that identifies a device from the unique properties of its GPU stack.
  Specifically, we show that variations in speed among the multiple execution units that comprise a GPU can serve as a reliable and robust device signature, which can be collected using unprivileged JavaScript.
  We investigate the accuracy of \DrawnApart under two scenarios.
  In the first scenario, our controlled experiments confirm that the technique is effective in distinguishing devices with similar hardware and software configurations, even when they are considered identical by current state-of-the-art fingerprinting algorithms.
  In the second scenario, we integrate a \textit{one-shot learning} version of our technique into a state-of-the-art browser fingerprint tracking algorithm.
  We verify our technique through a large-scale experiment involving data collected from over 2,500 crowd-sourced devices over a period of several months and show it provides a boost of up to 67\% to the median tracking duration, compared to the state-of-the-art method.

  \DrawnApart makes two contributions to the state of the art in browser fingerprinting.
  On the conceptual front, it is the first work that explores the manufacturing differences between identical GPUs and the first to exploit these differences in a privacy context.
  On the practical front, it demonstrates a robust technique for distinguishing between machines with identical hardware and software configurations, a technique that delivers practical accuracy gains in a realistic setting.
\end{abstract}

\section{Introduction}
Privacy is dignity. 
It is a human right.
In the domain of web browsing, the right to privacy should prevent websites from tracking user browsing activity without consent.
This is the case in particular for cross-site tracking, in which website owners collude to build browsing profiles spanning multiple websites over extended periods of time.
Unfortunately for users, the right to privacy conflicts with business interests.
Website owners are highly interested in tracking users for the purpose of showing them ads they are more likely to click on, or to recommend products they are more likely to purchase.

We focus on the common scenario where identifying a browser is equivalent to tracking a user.
The traditional way to track users is with cookies, small files that are stored by the browser at the request of the website, and forwarded to the website on demand~\cite{rfc2109}.
Recent regulations restrict and supervise the acquisition of private data by websites~\cite{GDPR,CCPA}, and in particular require that users consent to the use of cookies.
Furthermore, in an effort to protect users' privacy and curb tracking, modern browsers restrict cookie-based tracking, especially \textit{third-party trackers} that attempt to track users across multiple unrelated websites.

To overcome the limitations of cookies, less scrupulous websites often resort to an approach called \emph{browser fingerprinting}. 
To fingerprint a browser, the website provides a script that queries the browser's software and hardware configuration to collect attributes, such as the browser's version, OS, timezone, screen, language, list of fonts, or even the way the browser renders text and graphics.
The diversity of configurations allows websites to discriminate devices and, hence, to track users, without the use of cookies~\cite{DBLP:conf/sp/LaperdrixRB16}, even in a collection spanning millions of fingerprints~\cite{DBLP:conf/www/Gomez-BoixLB18}.
Surveying the Internet demonstrates that browser fingerprinting techniques are prevalent and used by many websites, no matter their category or ranking~\cite{DBLP:conf/sp/NikiforakisKJKPV13, DBLP:conf/ccs/EnglehardtN16, DBLP:conf/dimva/DureyLRR21}.

A significant difficulty of fingerprint-based tracking is that browser fingerprints evolve.
As shown by Vastel~\etal~\cite{DBLP:conf/sp/VastelLRR18}, fingerprints change frequently, sometimes multiple times per day, due to software updates and configuration changes.
To track a user, an adversary must link fingerprint evolutions into a single coherent chain.
This process is made difficult by the existence of devices with identical hardware and software configurations. 
It is difficult for an adversary to correctly link a fingerprint if there is a set of identical devices to which it might belong.
This limits the adversary's tracking duration.
In Vastel~\etal's evaluation over a dataset of nearly 100,000 fingerprints collected from 1,905 distinct browser instances, with a wide variety of fingerprinting attributes, their state-of-the-art machine learning technique was able to deliver a median tracking time of less than two months.

In this work, we bring a new insight to the challenge of browser fingerprinting identical computers, by observing that even nominally identical hardware devices have slight differences induced by their manufacturing process.
These manufacturing variations are shown to enable the extraction of unique and robust fingerprints from a variety of devices, both large and small, in other settings~\cite{DBLP:journals/pieee/HerderYKD14, DBLP:conf/dac/SuhD07}.
If an adversary were able to extract such a hardware fingerprint from the user's device, it would significantly extend the adversary's ability to track them. 
Extracting a hardware fingerprint from a browser, however, is far from trivial---since the attacker has little control. 
In particular, the attacker can only interact with the system through unprivileged JavaScript code and WebGL graphics primitives---the attacker has no control over the runtime environment of the system, including background processes and simultaneous user activity---and the attacker has very limited exposure to the system, making classical machine learning pipelines that rely on long training phases all but useless.  
Thus, in this paper we raise the following question:

\medskip
\begin{center}
\textit{Can browser fingerprinting work on devices with identical hardware and software configurations?}
\end{center}
\medskip

\parhead{Our Contribution.}
We claim this is possible, and we assess this claim with \DrawnApart, a technique that measures small differences among the \emph{Execution Units} (EUs) that make up a modern \emph{Graphics Processing Unit} (GPU).
By fingerprinting the GPU stack, \DrawnApart can tell apart devices with nominally identical configurations, both in the lab and in the wild. 
In a nutshell, to create a fingerprint, \DrawnApart generates a sequence of rendering tasks, each targeting different EUs.
It times each rendering task, creating a fingerprint trace.
This trace is transformed by a deep learning network into an \textit{embedding vector} that describes it succinctly and points the adversary towards the specific device that generated it.

We evaluate \DrawnApart in two main scenarios.
First, to validate the method's ability to distinguish nominally identical configurations, we perform a series of controlled experiments under lab conditions.
We experiment with multiple sets of identical devices from vendors including Intel, Apple, Nvidia and Samsung, and demonstrate that \DrawnApart consistently improves identification of these nominally identical devices, achieving high identification accuracy in multiple hardware configurations, even though state-of-the-art browser-based fingerprinting methods cannot tell them apart.
Second, to show that \DrawnApart affects user privacy, we integrate the technique into Vastel \etal's state-of-the-art fingerprinting algorithm from IEEE S\&P 2018~\cite{DBLP:conf/sp/VastelLRR18}, which uses machine learning to link browser fingerprint evolutions. 
We show that the median tracking duration is improved by up to 66.66\% once we add the \DrawnApart fingerprint. 

In summary, this paper makes the following contributions: 

\begin{itemize}[nosep]
  \item We design and implement \DrawnApart\footnote{The artifact accompanying this paper can be found at: \url{https://github.com/drawnapart/drawnapart}.}, a GPU fingerprinting technique based on the relative speed of EUs, that observes minute differences between GPUs (\cref{sec:context}).
  \item We investigate the performance of our fingerprinting technique with multiple sets of identical devices, demonstrating that it can tell apart devices with identical hardware and software configurations (\cref{s:labeval}).
  \item We integrate \DrawnApart into Vastel \etal's fingerprinting algorithm and show, through a large-scale crowd-sourced experiment with over 2,500 unique devices and almost 371,000 fingerprints, that \DrawnApart delivers considerable gains to the tracking accuracy of this state-of-the-art approach (\cref{s:generalization}).
  \item We suggest possible countermeasures against our fingerprinting technique, and discuss their advantages and drawbacks (\cref{sec:countermeasures}).
\end{itemize}

\section{Background}\label{sec:background}

\subsection{Browser Fingerprinting}
Mowery \etal~\cite{mowery2011fingerprinting} discuss fingerprinting on the Web.
As they state, fingerprinting can be applied constructively or destructively.
An example of constructive use of fingerprints would be to identify fraudulent users trying to log in while masquerading as legitimate users.
Browser fingerprinting can be used to detect bots~\cite{DBLP:conf/ccs/BurszteinMPT16, DBLP:conf/esorics/JonkerKV19, vastel20}, or support authentication, where the fingerprint is used in addition to a traditional authentication mechanism~\cite{DBLP:conf/acsac/AlacaO16, DBLP:conf/dimva/LaperdrixABN19}.
A destructive use might involve tracking users without consent~\cite{DBLP:conf/ccs/AcarEEJND14, DBLP:conf/ccs/EnglehardtN16}.
In this scenario, fingerprinting is used to augment or replace cookies---\eg to track across multiple domains, or when users disable or delete cookies.
Our technique can be applied to either scenario.

Many fingerprinting techniques exist in the wild~\cite{DBLP:conf/nordsec/BodaFGI11,DBLP:conf/pet/Eckersley10,mowery2012pixel,DBLP:journals/corr/NakiblySY15}.
They rely heavily on differences in devices' hardware and software characteristics found in HTTP header fields and JavaScript attributes.
The key challenge is to identify features and attributes that further discriminate devices and allow for their unique identification, and to overcome the tendency of these features to evolve over time because of changes to the user's software, configuration, or environment.

\subsection{GPU Programming}\label{sec:webgl}
The \emph{Graphics Processing Unit} (GPU) is specialized hardware for rendering graphics. 
GPUs have highly parallel architectures that are composed of multiple \emph{Execution Units} (EUs), or \emph{shader cores}, which can independently perform arithmetic and logic operations. 
Most consumer desktop and mobile processors from the past decade have on-chip GPUs with multiple EUs.
For example, the UHD Graphics 630 GPU---integrated into Intel Core\,i5-8500 CPUs---includes 24 EUs, while the Mali-G72 GPU---integrated into the Samsung Exynos 9810 chipset used in Galaxy S9, S9+, Note9, and Note10 Lite devices---includes 18 EUs.

\emph{Web Graphics Library} (WebGL) is a cross-platform API for rendering 3D graphics in the browser~\cite{webgl}.
WebGL is implemented in major browsers including Safari, Chrome, Edge, and Firefox.
Derived from native OpenGL~ES\,2.0, a library designed for developing graphic applications in C++, WebGL implements a JavaScript API for rendering graphics in an HTML5 canvas element.
WebGL takes a representation of 3D objects as a list of \emph{vertices} in space and information on how to render them, and translates them into a two-dimensional raster image that can be displayed on screen.
WebGL abstracts this process as a pipeline. 
Two pipeline steps which are of interest to this work are the \emph{vertex shader}, which places the vertices in the two-dimensional canvas, and the \emph{fragment shader}, which determines the color and other properties of each fragment.
The vertex and fragment shaders can run user-supplied programs, written in a C-derived programming language named \textit{GL Shading Language} (GLSL).

\section{GPU Fingerprinting}\label{sec:context}
\subsection{Motivation}
Similar to past work~\cite{DBLP:conf/pet/Eckersley10, DBLP:conf/sp/LaperdrixRB16}, we aim to uniquely identify devices.
However, unlike previous work, which rely on the diversity of hardware and software configurations, we focus on distinguishing identical devices.
As we show experimentally, this additional distinguishing power can considerably enhance the tracking capabilities of existing fingerprinting methods.
To do so, we incorporate techniques similar to the arbiter-based \emph{Physically Unclonable Function} (PUF) concept of Lee \etal~\cite{Lee04atechnique}.
In an arbiter PUF, the statistical delay variations of wires and transistors across multiple instances of the same integrated circuit design are used to uniquely identify individual instances of the integrated circuit.
In our case, we harness the statistical speed variations of individual EUs in the GPU to uniquely identify a complete system.

\subsection{Design}
With unfettered access to the GPU, an adversary could measure the speed of each EU and use those measurements as a fingerprint.
However, websites only have limited access to the GPU through the JavaScript and WebGL APIs. 
WebGL provides a high-level abstraction that makes it a challenge to target specific EUs and to time computations accurately.

We overcome this challenge by using short GLSL programs executed by the GPU as part of the vertex shader (cf. \cref{sec:webgl}).
We rely on the mostly predictable job allocation in the WebGL software stack to target specific EUs.
We observe that, when allocating a parallel set of vertex shader tasks, the WebGL stack tends to assign the tasks to different EUs in a non-randomized fashion.
This allows us to issue multiple commands that target the same EUs.
Finally, instead of measuring specific tasks, we ensure that the execution time of the targeted EU dominates the execution time of the whole pipeline.
We do so by assigning the non-targeted EUs a vertex shading program that is quick to complete, while assigning the targeted EUs tasks whose execution time is highly sensitive to the differences among individual EUs.
As shown in \cref{f:concept}, our fingerprint is created by executing a sequence of drawing operations.
We measure the time to draw a sequence of points with carefully chosen shader programs. 
The technique consists of three main steps:

\begin{figure}[t]
  \begin{centering}
      \includegraphics[width=.8\linewidth]{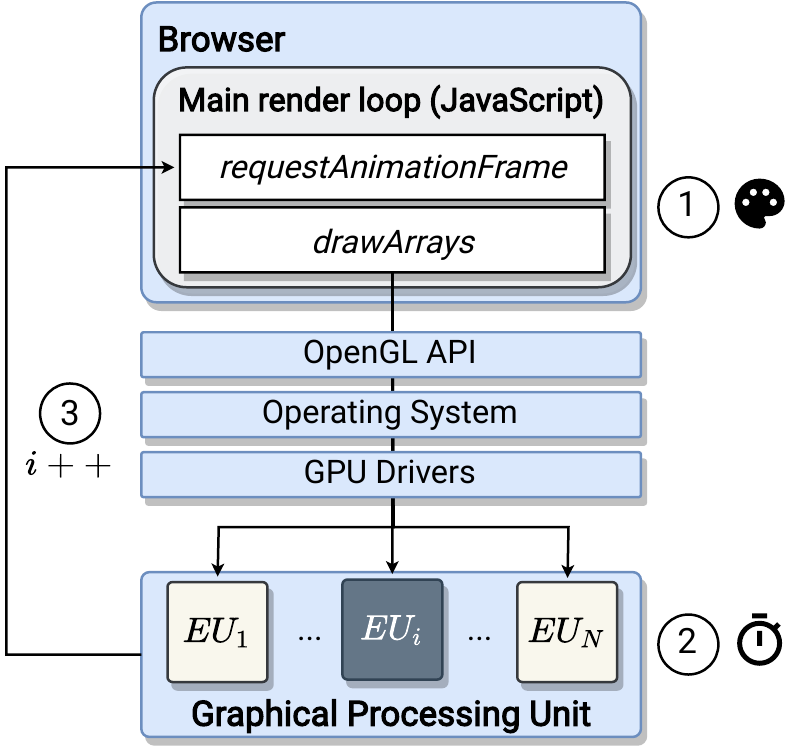}
  \par\end{centering}
  \caption{Overview of our GPU fingerprinting technique: (1) points are rendered in parallel using several EUs; (2) the EU drawing point $i$ executes a stall function (dark), while other EUs return a hard-coded value (light); (3) the execution time of each iteration is bounded by the slowest EU. }
  \label{f:concept}
\end{figure}

\parhead{Render.}
We instruct the WebGL API to draw a number of points in parallel. 
Points are the simplest object that WebGL can draw, and each consists of only a single vertex. 
Using points minimizes the noise from the pipeline and its interference with our technique.
The position of each point is determined by an attacker-controlled vertex shader.

\parhead{Stall.}
For most points, the attacker-controlled vertex shader returns a hard-coded value.
For a specific subset of the points the shader applies a function, which we call a \emph{stall function}, to compute the point's position.
The manner in which the entire graphics stack distributes the points to be drawn to the EUs allows us to influence which EU is chosen to run the stall function.
It takes much longer to compute the position with the stall function than the hard-coded value.
As a result, the time needed to render the entire set of points corresponds to the time taken by the EUs running the stall function.

\parhead{Trace Generation.}
We execute the drawing command several times, each time selecting a different vertex to stall.
For each execution, we store the time taken.
The fingerprint output by our technique is therefore a vector, named a \emph{trace}, which contains the sequence of timing measurements.

We note that prior browser fingerprinting techniques extract \textbf{deterministic} fingerprints, which remain identical as long as the device's software and configuration have not changed.
Our technique, in contrast, is based on timing measurements and, as such, is \textbf{non-deterministic}---multiple measurements made on the same device will return different values due to the effects of measurement noise, quantization, and the impact of other tasks running at the same time.

\subsection{Implementation}\label{web_implementation}

We now describe the implementation of each design step.

\parhead{Render.}
The WebGL API exposes the \texttt{drawArrays()} function, which allows dispatching multiple drawing operations in parallel to the GPU.
We invoke \texttt{drawArrays()} several times, each time rendering multiple points in parallel.
%
\Cref{fig:main_render} describes our main render loop. 
We execute the rendering process by calling \texttt{drawArrays} (line~\ref{line:draw_arrays}).
For each iteration, we save the time to execute \texttt{drawArrays} into the \texttt{trace} array.
We evaluated several ways of measuring the rendering time, as explained further in \cref{s:timers}.
Briefly put, the \textbf{onscreen} measurement method executes a relatively small number of computationally intensive operations, while the \textbf{offscreen} and \textbf{GPU} measurement methods execute a larger number of less computationally intensive operations. 
The full source code for these settings can be found in our artifact repository, as listed in \cref{sec:conclusion}.
After \texttt{point\_count} iterations, the code sends the \texttt{trace} array to our back-end server (line~\ref{line:send_fingerprint}), and terminates the loop.

\begin{figure}[htb]
\noindent\begin{minipage}[t]{\linewidth}
  \begin{lstlisting}[language=JavaScript,escapechar=|,caption={Main Render loop, onscreen setting (JavaScript).},label={fig:main_render}]
function render_loop() {
  if (point_index < point_count) {
      // Stall the current point
      gl.uniform1i(shader_stalled_point_id, point_index); |\label{line:set_uniform}|
      gl.drawArrays(gl.POINTS,0,point_count);|\label{line:draw_arrays}|
      // Save the rendering time
      var dt = performance.now() - prev_time;
      prev_time = performance.now();      |\label{line:measure_time}|
      trace.push(dt);
      // Prepare to stall the next point
      point_index++;
      requestAnimationFrame(render_loop); |\label{line:request_animation}|
  } else {
    // Finish and send the trace to the server
    send_trace();	                |\label{line:send_fingerprint}|
  }
}
\end{lstlisting}
\end{minipage}
\end{figure}

\parhead{Stall.}
In the current implementation of WebGL, a single call to \texttt{drawArrays()} generates multiple drawing operations in the underlying graphics API, which appear to assign vertices to EUs in a deterministic order during vertex processing.
The operations are differentiated by a global variable, named \texttt{gl\_VertexID}.
This special variable is an integer index for the current vertex, intrinsically generated by the hardware in all of the graphics APIs used to implement WebGL as it executes \texttt{gl.drawArrays}.
We created a vertex shader in GLSL that examines the \texttt{gl\_VertexID} identifier, and executes a computationally intensive \textit{stall function} only if it matches an input variable named \texttt{shader\_stalled\_point\_id} provided by the JavaScript code running on the CPU. 
\Cref{fig:shader_code} describes the vertex shader code.

 \begin{figure}[htb]
  \noindent\begin{minipage}[t]{\linewidth}
\begin{lstlisting}[language=GLSL,escapechar=|,caption={Vertex shader with stall function, onscreen setting (GLSL).},label={fig:shader_code}]
uniform int shader_stalled_point_id; |\label{line:shared_from_js}|
void main(void) {
  // Stall on this vertex?
  if(shader_stalled_point_id == gl_VertexID) {
    gl_Position = vec4(stall_func(),0, 1,1);  |\label{line:perform_stall}|
  } else { 
    gl_Position = vec4(0, 0, 1 ,1);
  }
  gl_PointSize = 1.0; |\label{line:mandatory_position}|
}

\end{lstlisting}
\end{minipage}
\end{figure}

In the onscreen setting, the vertex shader checks if \texttt{shader\_stalled\_point\_id} equals \texttt{gl\_VertexID}. 
In the offscreen and GPU settings, the vertex shader treats \texttt{shader\_stalled\_point\_id} as a bit mask and checks if bit \texttt{1 <{}< gl\_VertexID} is set. 
In both cases,  if the point is selected the vertex shader program executes the stall function (line~\ref{line:perform_stall}).
Otherwise, the shader exits quickly.

\parhead{Trace Generation.}
By executing this parallel drawing operation multiple times, each with a different value for \texttt{shader\_stalled\_point\_id}, we iterate over the different EUs and measure the relative performance of each.
The output is a trace of multiple timing measurements, corresponding to the time taken by the targeted EU to draw the scene. 

\subsection{Raw Traces}
\begin{figure}[tb]
  \includegraphics[width=\linewidth]{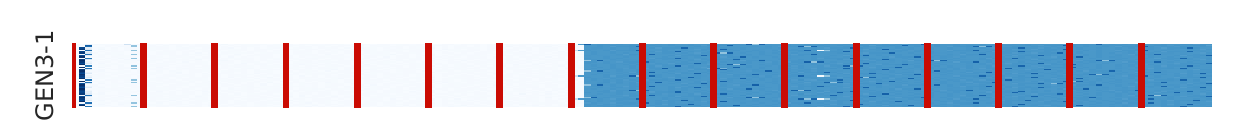}
  \includegraphics[width=\linewidth]{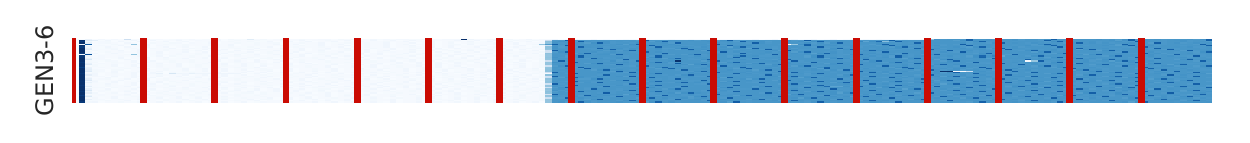}
  \caption{Raw traces from two different \GenIII devices.\label{f:raw}}
\end{figure}

Before evaluating \DrawnApart, we tested whether we can visually distinguish devices.
\cref{f:raw} shows traces collected from two \GenIII devices.
We collect 50 traces from each device, each trace consisting of 176 measurements of 16 points.
The measurements are divided into 16 groups of 11, where in each group we stall a different point.
The color of a point indicates the rendering time, ranging from virtually 0 (white) to ~90\,ms (blue).
Red vertical bars indicate group boundaries.
As we can see, the rendering time in the first half of the traces is significantly faster than in the second half.
Moreover, while there are some timing variations in the traces of the same device, the traces display patterns that are distinct between devices, 
allowing us to distinguish them.

\section{Evaluation Overview}\label{sec:results}

\subsection{Motivation}\label{sec:eval_motivation}

We claim that our new method provides a tangible advantage over deterministic GPU-based fingerprinting. 
To establish this claim, we evaluate our system in a lab setting and in the wild.

In the \textbf{lab setting}, we assume the attacker can collect training traces from a set of identical machines (same hardware and software), running under identical environmental conditions.
Next, the attacker is given a single trace and is tasked with identifying the machine that generated the trace.
Our primary metric of evaluation in this setting is the \textbf{accuracy gain}, which measures the multiplicative gain in accuracy of a classifier that incorporates our non-deterministic method, when compared to a classifier which only uses deterministic inputs. 
An accuracy gain of~1 means that the classifier provides no advantage over traditional methods, while higher values show that it gives the attacker an advantage. 
The lab setting provides the most advantageous conditions for our classifier, for several reasons.
First, existing deterministic schemes cannot tell apart identical devices, as we demonstrate experimentally, resulting in a very low base rate.
Second, the attacker can tailor the attack to the particular class of devices to be discriminated, and thus choose optimal parameters for the target hardware. 
Third, the workload on the target machines is controlled, minimizing measurement noise.
Finally, the attacker is not concerned with detectability or compatibility, and can run an experiment that takes a long time, that uses partially supported hardware features, or that is noticeable to the user.

We also evaluate our system \textbf{in the wild}.
More specifically, we evaluate how our method can be applied to track devices from a set of over 2,500 machines with 1,605 distinct GPU configurations, recruited through a crowd-sourcing experiment.
We first perform a standalone evaluation of our method, in the absence of additional identifying features.
We then provide additional deterministic features to the classifier, including the browser version, screen dimensions, HTTP headers, and other similar attributes. 
State-of-the-art fingerprinting techniques can produce unique browser fingerprints through the consideration of these signals, but these fingerprints are not ideal for tracking users since they evolve over time~\cite{DBLP:conf/sp/VastelLRR18}.
We therefore measure the added distinguishing power our method provides to existing browser fingerprinting schemes, with the primary metric of evaluation being the \textbf{additional tracking time} made possible through the combination of our novel technique with existing schemes.

The in-the-wild setting is more challenging.
First, the technique must perform well across a large number of devices, precluding tailored attacks, and the attacker is prohibited from using any trace collection method that is overly intrusive or time-consuming.
Second, the attacker's choice of machine learning pipelines is constrained. 
In particular, the attacker cannot use a long training phase since this does not make sense in the context of browser fingerprinting---the fingerprint should be useful at once, and not depend on the victim spending hours on the attacker's website.
The attacker must also be able to accommodate new devices joining the dataset in real-time, and should not be required to spend multiple CPU hours retraining the classifier every time a new device is detected.
Finally, the attacker cannot control the runtime characteristics of the machine being fingerprinted.
Our method will have to be tolerant to workload variations, GPU payloads from other tabs, browser and system restarts, and so on.

In the following section, we study the lab setting to demonstrate an upper bound on our classifier's potential accuracy gain, and to investigate parameter choices and their trade-offs on accuracy, compatibility and performance.
In \cref{s:generalization}, we select a single set of parameters and launch a large-scale crowd-sourced experiment in the wild, showing the advantage of our method in a realistic setting.

\subsection{Machine Learning Pipelines}\label{sec:classification}
We use two machine learning approaches to evaluate our fingerprinting technique. 
In the \textbf{lab setting}, we cast our fingerprinting problem as a conventional multinomial classification task, where the input is the trace of $N$ rendering times, and the output is the label of the device assumed to have generated this trace. 
We evaluated several classical machine learning models suitable for this task, including tree-based classifiers, $k$-Nearest Neighbors classifiers, Linear Discriminant Analysis, and Support Vector Machines. 
We ultimately chose to use the Random Forest ensemble classification algorithm~\cite{DBLP:journals/ml/Breiman01,liaw2002classification}, as it empirically delivered the best classification results in terms of accuracy.
We did not apply any feature engineering, submitting the raw traces into the classification algorithm.
To make sure we did not overfit our model, we applied a 5-fold train-test split to the data, and collected the mean accuracy reported by the folds, as well as the standard deviation among folds.

To evaluate our system \textbf{in the wild}, we needed a more elaborate pipeline for the reasons listed in \cref{sec:eval_motivation}.
Our method relies on \emph{neural networks} and consists of several steps:
\begin{enumerate*}
\item We preprocessed our traces by normalizing and reshaping them into matrix form.
\item We trained a convolutional neural network (\emph{CNN}) to solve the multinomial classification task.
\item We transformed the classification network into an embedding network using the semi-hard triplet loss algorithm of Schroff \etal~\cite{DBLP:conf/cvpr/SchroffKP15}.
\end{enumerate*}
The resulting network is capable of transforming our trace into a representation called an \emph{embedding}.
Because of the way the network is designed, the Euclidean distance between two traces from the same device will be small, while the Euclidean distance between traces from different devices will be large.
This allows the inference part of the classification to use the $k$-Nearest Neighbors classifier---given an unknown trace, measure the distance between its embedding and the embeddings of all known traces, and output the label of the embedding at the shortest distance.
The simplicity of this classifier means the adversary can add new devices to the dataset simply by recording a few new traces and without retraining the entire network, a desirable property known as \emph{few-shot learning}.

To ensure we did not overfit our in-the-wild model, we split our training dataset into two mutually exclusive parts, each with different labels, performed the evaluation on each part in isolation, and observed that the accuracies for each split were roughly the same.
More details about the training process and dataset splits can be found in \cref{s:generalization}.

\section{Lab Setting}\label{s:labeval}

\begin{table*}[htb]
  \centering
  \caption{Accuracy gains achieved under lab conditions}\label{t:lab_results}
\begin{tabular}{llllrrr}
\toprule
Device Type                                         & GPU                                     & \begin{tabular}[c]{@{}l@{}}Device \\ Count\end{tabular} & Timer     & \multicolumn{1}{c}{\begin{tabular}[c]{@{}c@{}}Base \\ Rate (\%)\end{tabular}} & Accuracy (\%)                & Gain \\ 
\midrule
\multirow{2}{*}{Intel i5-3470 (\GenIII Ivy Bridge)}   & \multirow{2}{*}{Intel HD Graphics 2500} & \multirow{2}{*}{10}                                     & Onscreen  & 10.0                                                                          & 93.0\pmtxt0.3 & 9.3  \\
                                                    &                                         &                                                         & Offscreen & 10.0                                                                          & 36.3\pmtxt1.6 & 3.6  \\\hline
\multirow{3}{*}{Intel i5-4590 (\GenIV Haswell)}      & \multirow{3}{*}{Intel HD Graphics 4600} & \multirow{3}{*}{23}                                     & Onscreen  & 4.3                                                                           & 32.7\pmtxt0.3 & 7.6  \\
                                                    &                                         &                                                         & Offscreen & 4.3                                                                           & 63.7\pmtxt0.6 & 14.7 \\
                                                    &                                         &                                                         & GPU       & 4.3                                                                           & 15.2\pmtxt0.5 & 3.5  \\\hline
\multirow{3}{*}{Intel i5-8500 (\GenVIII Coffee Lake)}  & \multirow{3}{*}{Intel UHD Graphics 630} & \multirow{3}{*}{15}                                     & Onscreen  & 6.7                                                                           & 42.2\pmtxt0.7 & 6.3  \\
                                                    &                                         &                                                         & Offscreen & 6.7                                                                           & 55.5\pmtxt0.8 & 8.3  \\
                                                    &                                         &                                                         & GPU       & 6.7                                                                           & 53.5\pmtxt0.8 & 8.0  \\\hline
\multirow{2}{*}{Intel i5-10500 (\GenX Comet Lake)} & \multirow{2}{*}{Nvidia GTX1650}         & \multirow{2}{*}{10}                                     & Offscreen & 10.0                                                                          & 70.0\pmtxt0.5 & 7.0  \\
                                                    &                                         &                                                         & GPU       & 10.0                                                                          & 95.8\pmtxt0.9 & 9.6  \\\hline
\multirow{2}{*}{Apple Mac mini M1}                  & \multirow{2}{*}{Apple M1}               & \multirow{2}{*}{4}                                      & Offscreen & 25.0                                                                          & 46.9\pmtxt0.4 & 1.9  \\
                                                    &                                         &                                                         & GPU       & 25.0                                                                          & 73.1\pmtxt0.7 & 2.9  \\\hline
Samsung Galaxy S8/S8+                               & Mali-G71 MP20                           & 6                                                       & Onscreen  & 16.7                                                                          & 36.7\pmtxt2.7 & 2.2  \\\hline
Samsung Galaxy S9/S9+                               & Mali-G72 MP18                           & 6                                                       & Onscreen  & 16.7                                                                          & 54.3\pmtxt5.5 & 3.3  \\\hline
Samsung Galaxy S10e/S10/S10+                        & Mali-G76 MP12                           & 8                                                       & Onscreen  & 12.5                                                                          & 54.1\pmtxt1.5 & 4.3  \\\hline
Samsung Galaxy S20/S20 Ultra                        & Mali-G77 MP11                           & 6                                                       & Onscreen  & 16.7                                                                          & 92.7\pmtxt1.8 & 5.6  \\
\bottomrule
\end{tabular}
\end{table*}

The objective of the lab setting is to discover \DrawnApart's highest accuracy, and assumes that the attacker customizes the attack to the class of device and ignores aspects of detectability, compatibility or performance.

\parhead{Evaluated Devices.} 
\Cref{t:lab_results} lists the devices used in the lab setting.
We used 88 devices from nine distinct hardware classes, including desktops and mobile devices.
The desktops include multiple generations of Intel processors, all running Windows 10, as well as a set of Apple Mac mini devices with an Apple M1 chip, running MacOS X Version 11.1.
Other than the \GenX devices, which had discrete Nvidia GTX1650 GPUs, all desktops used integrated graphics.
For each class, the devices were purchased through the same order, configured with our University's official operating system image, and located in the same temperature-controlled lab. 
The mobile devices include multiple generations of Samsung Galaxy devices, all sourced through the Samsung Remote Test Lab~\cite{Remotetestlab}. 
All the mobile devices were Android-based and featured Samsung Exynos CPUs and Mali GPUs.

\parhead{Comparison With Prior Fingerprinting Techniques.}
Before evaluating our technique, we reproduced and tested several state-of-the-art web-based fingerprinting techniques. 

\textbf{UniqueMachine}, presented by Cao \etal at NDSS 2017~\cite{DBLP:conf/ndss/CaoLW17}, 
collects a ``browser fingerprint'', with mutable properties such as window size and IP address, and a more permanent ``computer fingerprint''. 
The UniqueMachine website offers a demo that outputs both fingerprints as 32-character hashes. 
We collected the fingerprints of all of the computers in our \GenIII, \GenIV, \GenVIII, and \GenX corpora using UniqueMachine, and confirmed that all computers in the same corpus were assigned the same computer fingerprint. 
Interestingly, the \GenIV and \GenX PCs shared the same computer fingerprint despite having different hardware configurations. 

\textbf{Fingerprint JS (FPJS)} is a commercial API offering ``browser fingerprinting as a service''. 
The paid-for version, called FPJS Pro, claims to provide ``unparalleled accuracy, ease of use, and security''~\cite{fingerprintjs2}.
FPJS Pro outputs a 20-character hash. 
The website provides a demo of FPJS Pro. 
We collected the fingerprints of all computers in our \GenIII, \GenIV, \GenVIII, and \GenX corpora using the demo website.
In the \GenIII dataset, all but one computer had the same fingerprint. 
Similarly to UniqueMachine, all of the computers in the \GenIV and \GenX corpora had identical FPJS fingerprints. 
Finally, FPJS divided the \GenVIII corpus into three clusters: two clusters with seven computers each, and the final cluster with one computer.

\textbf{Clock around the Clock}, proposed by S{\'{a}}nchez{-}Rola \etal at CCS 2018~\cite{DBLP:conf/ccs/Sanchez-RolaSB18}, is an alternative to GPU-based fingerprinting.
This method is designed to exploit ``small, but measurable, differences in the clock frequency'' by measuring the precise execution times of a series of CPU-intensive operations.
To calculate the fingerprint, the computer invokes the cryptographic random number generator \texttt{crypto.getRandomValues} 1,000 times for 50 different input sizes, then generates a vector of the most common timing value, or mode, for each of the input sizes.
We reproduced the web-based variant of the method, and tested it on our \GenIV corpus.
We found that the modes did not contain any data useful for fingerprinting.
This is likely because since July 2018 Chrome contains countermeasures designed to prevent fine-grain timing measurements, as part of the wider fallout of the Spectre attacks~\cite{low_resolution_1, low_resolution_2, low_resolution_3, low_resolution_4}.
All our measurements returned either zero or five microseconds (with some added randomness).
We conclude that, currently, the method presented by S{\'{a}}nchez{-}Rola \etal is not practical.

\subsection{Tuning the Trace Parameters}\label{s:parameters}
We search for the parameter settings that provide the optimal accuracy gain for the different hardware configurations.

\parhead{Stall Function Operator Selection.}
Each model and generation of GPU has a different micro architecture.
For example, the third-generation Intel integrated GPU has a single arbiter, which dispatches tasks to all EUs, while fourth-generation GPUs adopt a hierarchical micro-architecture with multiple arbiters.
Intel GPUs also have \textit{Advanced Math (AM) Units}, which are tasked with executing less common operations such as trigonometric operations. 
The amount and location of these AM units differs among GPU generations, and even within different GPU types from the same generation.
The design of GPUs by Nvidia, ARM and Apple is obviously different as well.
We hypothesize that, due to these differences, the accuracy gain provided by our method will vary, depending both on the choice of stall functions and target hardware.
To test this, we evaluated a representative set of operators, including trigonometric operations, logical bit-wise operations, and general floating-point operations.
The set of operators selected can be found in~\cref{app:detailedeval}. 

\parhead{Timing Measurement Method.}\label{s:timers}
Scene rendering is performed in the GPU context, which is asynchronous to the CPU context.
Simply measuring the time it takes the CPU to execute the draw operation, for example by calling \performancenow immediately before and after the call, does not provide any usable insight about the GPU.
We therefore considered three measurement methods that are capable of measuring the actual drawing time of the GPU.

In the \textbf{onscreen} method, we render the scene to a standard HTML canvas element and then call \texttt{Window.requestAnimationFrame}.
This function is passed a callback function that is called after the rendering is complete. 
Timing information is then extracted from within the callback.
The onscreen method is the most compatible of those we evaluated, but browsers do not call \texttt{requestAnimationFrame} at a rate higher than the browser's maximum frame rate, which is typically 60 Hz.
Thus, using this method requires that each iteration of our rendering operation take at least 16\,ms to provide us with useful information. 
Even though the canvas element is on screen, it can be made zero-size or invisible via styling, making the fingerprinting operation invisible to the user.
Collecting the fingerprint does cause a noticeable slowdown for the user since it runs in the browser's main context.

In the \textbf{offscreen} method we use a worker thread and render the scene to an \texttt{OffscreenCanvas} object.
This does not affect the user's main context and does not slow down the user.
After rendering the scene, we call the \texttt{convertToBlob} method of the \texttt{OffscreenCanvas}, causing it to execute all instructions currently in the WebGL pipeline, and ultimately return a binary object representing the image contained in the canvas.
We measure the time it takes to execute this command.
Since there is no frame rate limit in this setting, each iteration of the rendering operation can take less time, allowing us to use  more iterations.
At the time of writing, \texttt{OffscreenCanvas} is supported on Chrome browsers, hidden behind a flag on Firefox, and partially supported in the Technical Preview build of Safari.

The \textbf{GPU} method is the third method we evaluate. 
It is a modification of the offscreen method that does not measure timing on the CPU side.
Instead, the WebGL disjoint timer query method is used to directly measure the duration of a set of graphics commands on the GPU side.
To perform this measurement, we call \texttt{beginQuery}, issue the drawing operations, and call \texttt{endQuery}. 
Using \texttt{getQueryParameter}, we retrieve the elapsed time on the GPU side.
This disjoint timer query command was previously used for side-channel attacks by Frigo \etal in their work in IEEE S\&P 2018~\cite{DBLP:conf/sp/FrigoGBR18}.
As a result, support for this timer was disabled in Chrome version 65.
However, with the introduction of Site Isolation~\cite{siteIsolation}, it was deemed safe to be re-enabled in Chrome version 70~\cite{webgl_timer_enable}.
In contrast to CPU-side timers, whose resolutions have been severely reduced to a few microseconds with jitter to mitigate against transient execution attacks~\cite{rokicki21timers}, the GPU-side timer offers microsecond resolution with no jitter even on the most modern versions of Chrome~\cite{chromeGpuMicro}.
This GPU-based timer thus has the potential to be the most accurate and the least sensitive to activity on the CPU side.
On the other hand, its accuracy varies dramatically between different GPU architectures, and it is not supported by the commonly used Google SwiftShader renderer.

\parhead{Number Of Points To Render.}
Our fingerprinting scheme relies on multiple iterations of a drawing command, where each iteration exercises a certain subset of the EUs while leaving the other EUs idle.
The number of iterations and the time each iteration takes to run will determine the total execution time.
However, it is reasonable to assume that capturing more data will provide better accuracy, and that relatively long workloads will mitigate the impact of the low-resolution timers available through JavaScript.
We ran two experiments to capture this trade-off.
The first was run in the onscreen setting, using the \GenIII corpus. 
The frame rate requirement of the onscreen setting limits each iteration to at least 16\,ms, as explained above.
The second experiment was run in the offscreen setting using the \GenIV corpus.
This setting allowed us to use much shorter workloads and to increase the number of iterations that can be run in a reasonable time period.
Thus, instead of assigning the stall function for each point only once per iteration, we tried all $2^n$ possible subsets of the set of points, allowing us to measure the \emph{contention} between EUs, as well as their individual speeds.

\subsection{Results}\label{s:lab_results}

\Cref{t:lab_results} summarizes the accuracy gains obtained in the lab setting using different timing methods.
The mobile devices were evaluated using the onscreen method only due to limited access to those devices.
\GenIII and \GenIV are not evaluated using the GPU timer method since their hardware does not support it.
All devices within each hardware class were sampled the same amount of times.
We observed that our Random Forest-based classifier approaches peak accuracy as the size of the training data set approaches 500 traces per label.
As the table shows, our scheme delivered significant accuracy gains, well above the base rate, in all scenarios, both for desktop and mobile devices.
The parameter choices, however, did affect the performance of our scheme.

\parhead{Effect Of Stall Function.} As expected, each of the operators we evaluated performed differently on the different hardware targets.
Specifically, in the onscreen setting, the \texttt{mul} operator delivered the best accuracy gains for the \GenIII and \GenIV corpora, while \texttt{exp2} was the best performer for the \GenVIII corpora, as described in more detail in \cref{app:detailedeval}.
The different mobile device corpora, which were also evaluated in the onscreen setting, also had different optimal operators: \texttt{pow} for Galaxy S8/S8+ and Galaxy S9/S9+, \texttt{atanh} for Galaxy S10e/S10/S10+ and \texttt{mul} for Galaxy S20/S20+/S20 Ultra.

In the offscreen setting, the \texttt{sinh} operator was consistently the best performer for the \GenIV and \GenVIII corpora, while \texttt{mul} was better than \texttt{sinh} for the \GenX corpora. 
We hypothesize that since the offscreen setting allowed us to trigger multiple execution units at the same time, and the amount of advanced math units that handle trigonometric operations is lower than the amount of EUs, the conflicts and race conditions that arise inside the GPU gave this operator additional discriminating power.

\parhead{Effect Of Timing Measurement Method.} As stated above, the offscreen method allowed us to execute more iterations than the onscreen method, allowing us to capture data about EU contention, as well as on the timing of individual EUs.
We were also interested in comparing the relative performance of the offscreen method, which measured time on the CPU side, and the GPU method, which used disjoint timer queries to measure performance on the GPU side.
We hypothesizes that the GPU method would be superior to the offscreen method, since the GPU-side timer has higher accuracy than the CPU-side timer, and is not affected by the timing jitter introduced by inter-process communications (IPC) between the GPU and the CPU.
In practice, we discovered that this is not always the case.
As shown in \cref{t:lab_results}, the GPU timer is better than the CPU timer for the Intel \GenX and Apple M1 corpora, has equivalent accuracy to the CPU timer on the \GenVIII corpus, and is actually less accurate than the CPU timer on the Intel \GenIV corpus. 
To make matters worse, the disjoint timer query WebGL extension is not supported on several popular WebGL stacks, most significantly the software-based Google SwiftShader.
Thus, the GPU-based timer is not appropriate for use in a large-scale experiment where the hardware configuration is not known beforehand.

\begin{figure}[!ht]
	\centering
	\includegraphics[width=\linewidth]{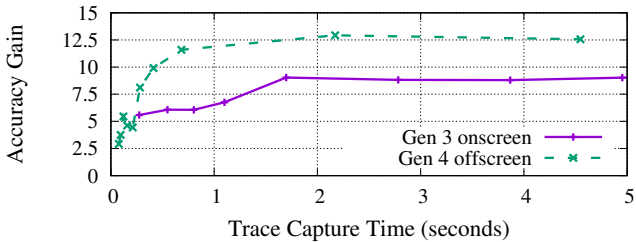}
	\caption{Accuracy gain as a function of trace capture time}
	\label{f:accuracy_vs_points_and_time}
\end{figure}

\parhead{Accuracy vs. Capture Time.} 
\Cref{f:accuracy_vs_points_and_time} shows the accuracy gain as a function of trace capture time, both for the \GenIII corpus using the onscreen collection method, and for the \GenIV using the offscreen collection method.
As the Figure shows, the accuracy gain of both methods approaches its optimal point when samples are collected for around 2 seconds. 
This is reached after about 80 iterations in the onscreen method and 1024 iterations in the offscreen method.

\parhead{Swapping Hardware.}
To reinforce our claim that the classification results are due to differences in the behavior of the GPUs, and not due to some residual differences among the computers, we selected two \GenIII computers, physically swapped their hard drives, and re-ran the fingerprinting classifier.
As expected, the fingerprinting classifier was not misled by the hard disk transplant, and was still able to label each of the two computers according to their CPU.
Next, we returned the hard drives to their original locations, and physically swapped the CPUs with integrated graphics of the two systems. 
As expected, the classifier followed the transplanted CPU, even though all other hardware was unmodified.

\subsection{Evaluation on Additional Browsers.}\label{s:more_browsers_lab}
We collected and evaluated traces from 16 devices from the \GenIV corpora using multiple additional browsers: Brave browser~\cite{Brave} (version \textsf{81.0.4044.113}), Edge~\cite{Edge} (version \textsf{96.0.1054.43}), Opera~\cite{Opera} (version \textsf{82.0.4227.23}) and Yandex browser~\cite{Yandex} (version \textsf{21.11.3.927}), all using the \textit{offscreen} method.
The accuracy showed a significant improvement over the base rate, which lies at $6.25\%$, with Edge, Brave, Opera and Yandex, delivering accuracies of 34.6\pmtxt0.6\%, 31.0\pmtxt0.3\%, 31.6\pmtxt0.7\%, and 31.1\pmtxt0.3\%, respectively.

We evaluated the stability of \DrawnApart over 21 devices of the \GenIV corpora for an extended period of time. 
We collect data for both Chrome and Firefox.
For Chrome, we use the \textit{onscreen} and \textit{offscreen} methods.
For Firefox, which does not currently support the \textit{offscreen} method, we are limited to the \textit{onscreen} method. 
We also chose to stall the EU for twice as many operations under Firefox, compared to Chrome, to account for the lower timer resolution found in Firefox.

\begin{figure}[htb]
	\centering
	\includegraphics[width=\linewidth]{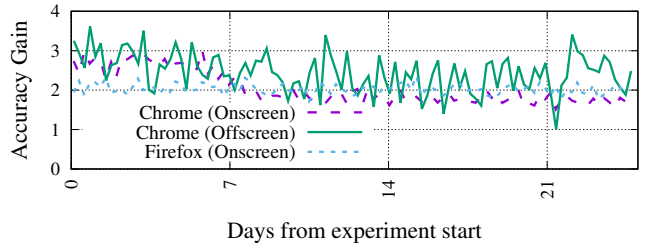}
	\caption{Additional Browsers -- Lab Evaluation}
	\label{f:more_browsers_lab}
\end{figure}

For 24 days, we repeatedly launched the browser, collected traces for 20 minutes  using the \textit{offscreen} method and for 40 minutes using the \textit{onscreen} method, then quit the browser and idled for 4 hours.
The first 4 cycles were used to train the Random Forest classifier, while the remaining cycles over the experiment's 24 days were used to evaluate its performance.
The results are summarized in~\Cref{f:more_browsers_lab} and show the accuracy to be above the base rate for each point in time. 
We observed that the \textit{offscreen} method yields slightly higher accuracy than \textit{onscreen}, and that the accuracy of both methods on Chrome slightly decay over time, while the accuracy of the \textit{onscreen} method on Firefox remains stable.
Finally, the accuracy in this experiment is lower compared to the results reported in~\Cref{t:lab_results}.
It is possibly due to repeatedly restarting the browser over the course of the experiment, as we discuss in~\Cref{s:limitations_and_observations}.   

\subsection{Summary}\label{s:lab-summary} 

Our results show that \DrawnApart can tell apart identical computers in a controlled lab setting.
Our next objective was to a realistic setting, in which the attacker has less control over the devices to be fingerprinted.
We did so by first evaluating \DrawnApart in a standalone setting, and then integrating it with a state-of-the-art browser fingerprinting algorithm.

\section{In-the-Wild Setting}\label{s:generalization}
Performing browser fingerprinting in the wild presents different challenges compared to what we experienced with the lab setting:
\begin{enumerate*}
\item The lab evaluation assumed a closed list of devices.
In the real world, new devices can be added at any time during the collection period, but we cannot re-train the model whenever it happens.
\item The lab evaluation assumed we had a long time to collect data and train over the devices.
In the real world, we do not have unlimited access to a device so the collection of data must be fast.
\item Finally, the lab evaluation assumed the devices were idle and in a controlled environment.
In the real world, we have to contend with variable computing loads, restarts, and updates to both the browser and the operating system.
\end{enumerate*}
In order to understand the potential impact of \DrawnApart in the real world, we collected $370,392$ traces from $2,550$ devices over 7 months and performed the two following evaluations:
\begin{itemize}
\item \textbf{Standalone evaluation}: Considering only \DrawnApart traces without any other information, we aim to see how our method performs at reidentifying a device among others.
In \cref{s:onshotlearningpip}, we propose a one-shot learning pipeline whose aim is to match a new trace with another known trace present in our dataset.
\item \textbf{Tracking over time}: Browser fingerprints evolve~\cite{DBLP:conf/pet/Eckersley10}. 
Vastel~\etal{} developed two algorithms to track evolutions and link fingerprints that belong to the same device~\cite{DBLP:conf/uss/VastelLRR18}.
In \cref{s:improving}, we show how \DrawnApart can improve the \FPStalker{} algorithms, which are the current \sota{} tracking algorithms, by increasing the duration users can be tracked.
Our main metric to evaluate the gain of our technique will be the \textbf{median tracking time}.
Contrary to the standalone evaluation, we use all the attributes listed in \cref{app:attributes} as well as the \DrawnApart traces.
\end{itemize}

\subsection{Dataset constitution}\label{s:dataset}

\parhead{Large-scale Experiment.}
To show \DrawnApart's practical advantages over traditional deterministic fingerprinting methods as used in \FPStalker{}, we launched a large-scale experiment with diverse hardware and software.
We integrated our \DrawnApart technique into the Chrome browser extension from the \AmIUnique crowd-source experiment~\cite{DBLP:conf/sp/LaperdrixRB16}.
The extension periodically collects the browser fingerprints of thousands of volunteers, allowing us to track their evolution.

\parhead{\DrawnApart Collection Parameters.}
The crowd-sourced experiment constrained our choices. 
Most importantly, we wanted to be as non-intrusive as possible, as to not cause any user-perceivable slowdowns.
In addition, we wanted to be compatible with various rendering stacks we encounter in the wild.
Finally, we were interested in selecting a stall function that discriminates a wide variety of hardware.
With these constraints, we selected the \textbf{offscreen} timing method, which is supported by all desktop versions of Chrome.
The onscreen method was not selected as it causes slowdowns, and the GPU method was not selected since it is not supported by the Google SwiftShader renderer.
We chose the \texttt{sinh} stall function operator, which provided good performance during our tests.
We render all possible subsets of 10 points in each trace, for a total of $2^{10}=1,024$ iterations per trace.
This fingerprint takes a median time of 1.6 seconds to run.
It is collected by the extension using a worker thread, without affecting the user's interactions with the browser.
To increase our trace count, we repeated each collection seven times, for a median total run time of approximately 12 seconds.
We collected the traces every four hours. 

\parhead{Dataset Preparation.}
Our dataset contains $370,392$ fingerprints from $2,550$ unique devices.
In each fingerprint, we collect the attributes listed in \cref{app:attributes}, together with 7 \DrawnApart traces. 
We identify devices with the same GPU by looking at the \textit{WebGL renderer string} property. 
Over 90\% of the devices shared a renderer string with at least one additional device.
The largest observed group with same renderer string consisted of $534$ unique devices.

We split our dataset into three subsets, divided by measurement time:
$\oneMP$ contains $109,375$ samples collected between 3-Jan-2021 and 7-Feb-2021,
$\twoMP$ contains $46,293$ samples collected between 7-Feb-2021 and 31-Mar-2021, and
$\threeMP$ contains $214,724$ samples collected from 3-May-2021 to 8-Jul-2021.
We randomly choose 65\% of the devices in $\oneMP$ that have more than 28 samples, and refer to this subset as $\oneMP_{65}$.
The rest of $\oneMP$ will be referred to as $\oneMP_{\rest}$
The limit of 28 samples, or 196 \DrawnApart traces, was chosen to make sure the neural network will generalize well, by preventing it from overfitting on a small amount of traces of a specific device.
We normalized each trace and reshaped a vector of length 1024 into a 32x32 matrix.

\subsection{Standalone evaluation} \label{s:onshotlearningpip}
Before integrating our model with \FPStalker{}, we first evaluate it in isolation using only \DrawnApart traces and ignoring the other attributes.
In contrast to the classical ML model used in the lab setting, we used a neural network pipeline for the in-the-wild setting.
The ultimate goal of the pipeline is to generate quality embeddings in Euclidean space, which express the distance between traces.
We begin the process by creating a \emph{Convolutional Neural Network} (CNN)-based multinomial classifier.
The structure we selected for the classifier is inspired by Picek~\cite{DBLP:conf/space/Picek19}, and includes 
$N$ convolution blocks followed by a flatten layer, a dense layer, another L2-normalized dense layer without activation, and concluding with a fully connected layer with softmax activation.
Each convolution block contains a convolution layer, a dropout layer, and an average pooling layer.
We used scikit-optimize's Bayesian optimization~\cite{scikitoptimize} to search for the best parameters, as described in \cref{a:hyperparameters}, using 80\% of the traces in $\oneMP_{65}$ for training, and the remainder of $\oneMP_{65}$ for validation.
The parameter search took 48 hours on a server with four NVIDIA GEFORCE RTX 2080 Ti GPUs, two Intel Xeon Silver 4110 CPUs, and 128 GiB of RAM. 
The run yielded 79 valid neural networks. 
The best network achieved a training accuracy of 35.57\% and a validation accuracy of 33.82\%.

\parhead{Semi-Hard Triplet Loss Model.}
The next step in our ML pipeline is the transformation of the multinomial classifier into an embedding, using the triplet loss method.
Triplet loss minimizes the distance between an anchor and a positive, both of which have the same label, and
maximizes the distance between the anchor and a negative of a different label.
Semi-hard triplet loss means that we only use triplets that have a negative that is farther from the anchor than the positive, but still produces a positive loss~\cite{DBLP:conf/cvpr/SchroffKP15}.
We took our trained classification model, removed its last layer, and trained it again for 30 epochs on the same dataset as before, this time with a bigger batch size of 1024 preprocessed traces and with semi-hard triplet loss.
Batch size is important to the triplet mining process since we need sufficient examples in the batch to find enough semi-hard triplets.
We took the weights of the epoch that yielded a model with the best accuracy using a 1-Nearest Neighbor classifier.
The end-product of this process is a model that accepts preprocessed \DrawnApart traces as input and produces embeddings in a Euclidean space.
Labels are not involved in this process---we can take any \DrawnApart trace, even from a device that the model was not trained on,  feed it into the triplet loss model, and get Euclidean space embeddings.
We note that we optimized for the accuracy of the classification model, instead of the 1-Nearest Neighbor, to reduce the running time of our parameter search. 

\parhead{Evaluating The Classifier.} \label{oneshotpipeval}
The use of embeddings mandates using a $k$-Nearest Neighbors classifier for analyzing the outputs of the network.
Our metric for evaluation is the top-$k$ accuracy, which stands for the probability that the correct answer is one of the $k$ nearest neighbors of the selected trace, for $k=1$, 5, and 10, according to the distance metric output by the model.
\parhead{Base Rate Calculation.} The accuracy of a classifier should be compared to the \textit{base rate} obtained by a naive classifier with no access to the features.
In the case of a classical learning problem, the naive classifier can observe the training data and learn the apriori probabilities of each label.
Then, to get the best accuracy, this naive classifier will output the label of the most commonly observed device, or the $n$ most commonly observed devices for a top-$n$ setting.
The base rate in that case is therefore the cumulative proportion of these devices in the dataset.
In the case of a $k$-shot learning problem, the classifier does not know the apriori probabilities of each label, since it gets an equal amount of training data for each label.
The naive classifier in this case will just output a random label, or $n$ random labels for a top-$n$ setting.
The base rate in that case is only $n*(\#devices)^{-1}$.

\parhead{Train-Test Split Evaluation.} We evaluated our model in two ways: random train-test split, and $k$-shot learning.
In the train-test split evaluation, we randomly split each of the $\oneMP_{65}$, $\oneMP_{\rest}$ and $\twoMP$ datasets into two parts, using 80\% for memorizing and 20\% for testing.
We first used $\oneMP_{65}$ for evaluation. 
On this subset, the base rate is 1.00\% for top-1 accuracy, 3.51\% for top-5 accuracy and 6.15\% for top-10 accuracy.
To show that our network can generalize and work on traces it has never seen before, we next considered the performance of the network on $\oneMP_{\rest}$. 
On this subset, the base rate is 1.22\% for top-1 accuracy, 4.42\% for top-5 accuracy and 7.2\% for top-10 accuracy.
To show that our network generalizes to more devices and new traces, we evaluate it on $\twoMP$. 
$\twoMP$ contains devices from $\oneMP$, meaning that
the neural network was trained on some of the devices in $\twoMP$, but not all of them, but it was never trained on any traces from $\twoMP$.
On this subset, the base rate is 0.64\% for top-1 accuracy, 2.78\% for top-5 accuracy and 4.38\% for top-10 accuracy.
The results in \cref{t:wild_eval} demonstrate that our model accuracies are significantly better than the base rate for all of the three datasets.
The accuracies on $\oneMP_{65}$ and $\oneMP_{rest}$ datasets are roughly the same, showing the model responds well to new devices.
The small drop in the accuracy of $\twoMP$ despite a base rate of approximately half the other datasets, the addition of more devices and new traces and being collected at a later date, shows the model has generalized well.

\parhead{$k$-shot Learning Evaluation.} The $k$-shot learning evaluation was performed on the $\twoMP$ dataset.
We chose $\twoMP$ to evaluate $k$-shot learning because we used the traces from $\oneMP_{65}$ to train our triplet loss model, which would bias the results.
While some of the devices in this subset also appear in $\oneMP$, none of the traces in $\twoMP$ were used to train or validate the neural network.
In the memorizing phase, we memorize the first $k$ collections ($k\times 7$ \DrawnApart traces) of each device in $\twoMP$.
The rest of the traces of $\twoMP$ are used in the testing phase, again using a $k$-Nearest Neighbors classifier.
This is an evaluation that is close to real-world use.
An attacker would like to identify users with as few collections as possible.
This evaluation is harder than the previous one due to the small amount of data available for the memorizing phase.
In addition, the time difference between $\oneMP$ and $\twoMP$ requires the network to deal with concept drift. 
As mentioned above, the base rate in this setting is very small, because the attacker cannot learn anything about the distribution of the devices in the test set.
The results can be found in \cref{t:wild_eval}. 
As expected, they show a decrease in accuracy compared to the evaluation using random split, but our model still delivers significant accuracy beyond the base rate.
We thus conclude that \DrawnApart can be used for few-shot learning.

We leave the $\threeMP$ dataset to be used in the evaluation process of \FPStalker{} to test the model on a truly unseen dataset that reproduces in-the-wild conditions.

\begin{table}[tb]
	\centering
  	\caption{Standalone Performance of \DrawnApart In the Wild Using The Random Split (RS) and $k$-shot Methods} 
  	\label{t:wild_eval}
	  \begin{tabular}{@{}l@{}rrr@{}}
		\toprule
	    \textbf{Evaluation Method\hskip -0.1in}& \multicolumn{3}{c}{Accuracy (\textit{Base rate})} \\ (Dataset) & \multicolumn{1}{c}{\textbf{Top-1}} 	& \multicolumn{1}{c}{\textbf{Top-5}} 	& \multicolumn{1}{c}{\textbf{Top-10}}\\
        \midrule
		RS ($\oneMP_{65}$)     	 & 28.88\% (\textit{1.00\%})		& 56.36\% (\textit{3.51\%}) 			  & 68.70\% (\textit{6.15\%})                      \\
		RS  ($\oneMP_{rest}$)		 & 28.28\% (\textit{1.22\%})    & 55.09\% (\textit{4.42\%})  			  & 67.15\% (\textit{7.20\%})                       \\
    RS  ($\twoMP$)           & 23.33\% (\textit{0.64\%})    & 47.23\% (\textit{2.78\%})         & 58.83\% (\textit{4.38\%})                        \\
		1-Shot ($\twoMP$)        & 5.44\% (\textit{0.05\%})  		& 14.10\% (\textit{0.26\%})			    & 19.95\% (\textit{0.51\%})	                       \\
		5-Shot ($\twoMP$)        & 7.11\% (\textit{0.05\%}) 	  & 19.34\% (\textit{0.26\%})				  & 26.75\%  (\textit{0.51\%})                      \\
		10-Shot ($\twoMP$)       & 9.22\% (\textit{0.05\%}) 		& 22.77\% (\textit{0.26\%})	        & 31.09\%  (\textit{0.51\%})                      \\
		\bottomrule
	\end{tabular}
\end{table}

\begin{figure}
	\centering
	\includegraphics[width=\linewidth]{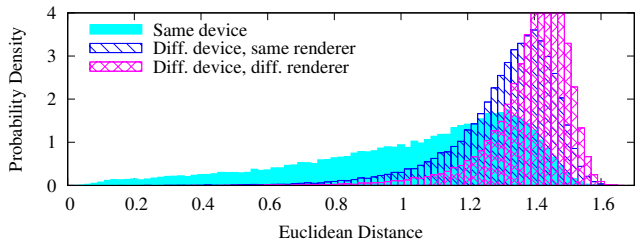}
	\caption{Performance of the \DrawnApart embedding function. A Euclidean distance below 0.65 indicates that the traces are likely to be from the same device.}
	\label{f:Euclidean_distances}
\end{figure}

\parhead{Visualizing Euclidean Distances.}
To visualize the performance of our few-shot learning pipeline, we computed the Euclidean distances between pairs randomly sampled from $\twoMP$ from the three following populations: Embeddings from the same device, embeddings from different devices that share the same renderer string, and finally embeddings from different devices with different renderer strings.
To eliminate correlations between traces in the same collection, we used only the first trace in the collections that we sampled from.
It means that we measured the distance between traces from different collections only.
\cref{f:Euclidean_distances} presents the probability density of the different distributions.
As the figure shows, embeddings from the same device get a lower Euclidean distance compared to embeddings from different devices, even if the device has the same GPU.
Of interest is that embeddings from different devices that share the same renderer string have a lower Euclidean distance compared to different devices that do not share the same renderer string. 
This confirms that \DrawnApart indeed fingerprints the GPU stack or an element correlated with the GPU stack.
We can also observe that if two traces have a Euclidean distance of less than 0.65, we can be almost certain that both traces came from the same device.
This is a strong property, we use it in the next section to improve \FPStalker{}.

\subsection{Evaluation on additional browsers in the wild.} While approximately 93.8\% of the traces found in our in-the-wild dataset $\twoMP$ come from users running the Google Chrome browser, some users submitted traces using other Chromium-based browsers.
We isolated non-Chrome users by filtering the traces according to their \textit{user-agent}, and analyzed the effectiveness of our standalone machine learning pipeline on these browsers as well.
The non-Chrome traces came from users running Edge, Opera and Yandex, which represented 5\%, 0.7\% and 0.5\% of the traces respectively.
We run the evaluation pipeline described in \cref{s:onshotlearningpip} for each browser, independently.
Our results show that the standalone pipeline's accuracy for Edge, Opera and Yandex is 52.6\%, 79.3\%, and 89.7\%, respectively.
The smaller amount of traces in this subset of the data results in a higher base rate when compared to the entire $\twoMP$ dataset---3\% for Edge, 17.9\% for Opera, and 27.6\% for Yandex.
These results, with the lab setting results, indicate that our fingerprinting technique identifies browsers from multiple vendors.
More details can be found in \cref{app:addl-browsers-in-the-wild}.

\parhead{Summary.} \label{summaryStandalone}
The results of the standalone evaluation, as summarized in~\Cref{t:wild_eval}, show a significant improvement over the base rate, demonstrating that \DrawnApart is effective on its own. 
However, it can be observed that the classifier's effectiveness is significantly reduced in the $k$-shot model, where the attacker has a limited trace budget to be used for training.
Putting these numbers into context is important.
In the world of browser fingerprinting, no single attribute differentiates all devices. 
While some attributes are more discriminating than others, it is their combination that is key to differentiating one device from another.
The standalone evaluation of \DrawnApart shows that our method has the potential to significantly contribute to fingerprinting accuracy.
In the following subsection, we empirically measure this contribution by using our method in conjunction with additional fingerprinting attributes.

\subsection{Integrating \DrawnApart with \FPStalker} \label{s:improving}
\FPStalker{} is the \sota{} fingerprint linking algorithm~\cite{DBLP:conf/sp/VastelLRR18}.
In this section, we show that \DrawnApart can be used to improve the \sota{}.

\parhead{Hybrid Algorithm.} \FPStalker{} has two distinct algorithms: one entirely rule-based, while the other combines rule-based constraints and machine-learning.
Vastel~\etal{} demonstrated that their hybrid variant of yielded better results on their dataset, but was slower than its rule-based counterpart.
As we are trying to prove the effectiveness of \DrawnApart in a real-world scenario, we chose to implement and optimize the hybrid \FPStalker{} algorithm, regardless of its speed.

\FPStalker{} consists in:
\begin{enumerate*}
\item a preprocessing step that discards fingerprints that contain inconsistencies or have been spoofed and cannot be normally found in the wild,
\item a training phase, in which the Random Forest algorithm is trained on a balanced dataset,
\item an inference phase, in which the trained model, combined with rules, compares incoming fingerprints to a pool of previously classified fingerprints and attempts to link them.
\end{enumerate*}
\cref{app:hybridfpwithda} lists the linking algorithm.

\parhead{Improving The Algorithm.} 
As mentioned in \cref{s:onshotlearningpip}, the output of the embedding network consists of 256 L2-normalized points that allow us to use a Euclidean distance to compute the similarity between embeddings.
\cref{f:Euclidean_distances} shows that the Euclidean distance is efficient, to an extent, in differentiating devices.
Based on the results obtained in \cref{oneshotpipeval}, which show that \DrawnApart can correctly classify devices with an acceptable accuracy, we decided to introduce the use of the generated embeddings as a complement to the machine-learning side of \FPStalker{}. 
We note that the results of our nude \FPStalker{} cannot be fully compared to the results obtained by Vastel~\etal{} for two main reasons:
\begin{enumerate*}
  \item Their dataset spans for longer than the dataset we use in our experiments.
  \item Flash-related attributes no longer exist,\cite{adobeendoflife}, impacting \FPStalker's effectiveness. 
\end{enumerate*}

Integrating \DrawnApart{} as a complement to \FPStalker{}'s machine-learning model is motivated by the fact that \FPStalker{} uses a series of conditions on the output of the Random Forest that makes its decisions too restrictive.
\FPStalker{}'s original code includes a function to optimize the threshold used by the Random Forest, which we adapted and ran on our dataset.
The resulting threshold yielded similar results, consequently comforting our observation that the rules associated to the output of the Random Forest are too restrictive, and discard too many fingerprints coming from the same browser instance.
On the other side, \cref{f:Euclidean_distances} shows that even though the Euclidean distances can be used to efficiently differentiate devices with a relatively low threshold, its usage alone may yield an unacceptable rate of false linkages due to a little percentage of different devices having low Euclidean distances.
To use \DrawnApart embeddings in \FPStalker{}, we average the seven embeddings that are collected with each fingerprint and we output an average embedding.
We used the previously generated averaged embeddings to compute the cosine similarity of the two compared fingerprints.
The resulting similarity is compared to a threshold we chose based on an analysis on the train dataset. This process is explained in the next paragraphs.
If the similarity of the two embeddings is above the chosen threshold, we classify the fingerprint as similar to the one being compared without further steps. 
The algorithm with the \DrawnApart additions, is available in \cref{app:attributes}.

\parhead{Choosing The Epsilon Threshold.} 
We chose the threshold by performing an analysis over similar and different devices on the train dataset.
We generate an equally balanced dataset from the training set comprising the cosine similarity of similar devices and different devices, and compare different percentiles of the distance of each group. 
As opposed to the Euclidean distance used in \cref{f:Euclidean_distances}, was chose the cosine similarity for \FPStalker because it is bounded by a more natural interval of $[-1; 1]$.
Our experiments showed that our threshold on the cosine similarity yielded better results than our Euclidean distance threshold.
Following our analysis, we noticed that the 5th-percentile of similar devices are all comprised below a similarity of $0.10$.
Consequently, we chose a threshold of $0.15$ in our experiments to account for a safety margin.

\begin{table}[t]
  \centering
  \caption{Average tracking time by collection period}
  \label{tab:trackingtimebycollectfrequency}
  \begin{tabular}{llll}
    \toprule
		\multirow{2}{*}{\textbf{Collection Period}}	& \multicolumn{2}{c}{\textbf{Tracking duration in days}} & \multirow{2}{*}{\textbf{Improvement}} \\
  												& \textbf{Nude FPS} 	& \textbf{FPS+DA}	\\
		\midrule
  2 days             & 17       & 26     & +52.94\%     \\
  3 days             & 17.25    & 25.5   & +47.82\%    \\
  4 days             & 17       & 28     & +64.70\%    \\
  5 days             & 17.5     & 27.5   & +57.14\%    \\
  6 days             & 18       & 30     & +66.66\%       \\
  7 days             & 17.5     & 28     & +60.00\%     \\
  \bottomrule
  \end{tabular}
  
\end{table}

\begin{figure}[t]
	\centering
	\includegraphics[width=\linewidth]{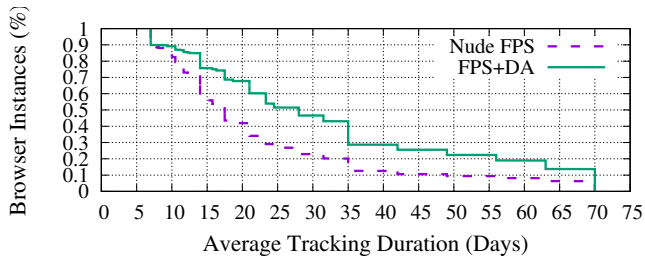}
	\caption{Differences in Average Tracking Time between \FPStalker{} (Nude FPS) and \FPStalker{} with \DrawnApart(FPS+DA)}
	\label{f:fps_avg_tracking}
\end{figure}

\parhead{Results.} 
We executed our revisited \FPStalker{} with its \DrawnApart{} addition on the dataset described in \cref{s:dataset}.
We first trained the Random Forest model on fingerprints in the $\oneMP$ subset.
We then executed the lambda optimization in order to run \FPStalker{} with its optimal parameters, as required by the original paper.
Finally, we executed the inference phase on $\threeMP$, which is unseen by the training phase of both \FPStalker{} and the embedding's network.
We execute both \FPStalker{} without our contribution, and our revisited version with \DrawnApart, on the same dataset for collection periods ranging from two to seven days.
\cref{tab:trackingtimebycollectfrequency} presents the average tracking duration obtained for each collection period, with a top improvement of 66.66\% compared to the original \FPStalker{} on a collection period of six days.
\cref{f:fps_avg_tracking} presents the average tracking duration with a collection period of seven days, as presented in the original paper, which represents tracking a user who visits a website once a week.
As the figure shows, adding \DrawnApart to \FPStalker{} increases the tracking time, raising the median average tracking time by \textbf{10.5 days}, from \textbf{17.5 days} to \textbf{28 days}.
This is a substantial improvement to stateless tracking, obtained through the use of our new fingerprinting method, without making any changes to the permission model or runtime assumptions of the browser fingerprinting adversary.
We believe it raises practical concerns about the privacy of users being subjected to fingerprinting.

\section{Discussion}\label{sec:discussion}

\subsection{Ethical Concerns}\label{s:ethics}
We integrated our fingerprinting algorithm into the Chrome browser extension from the \AmIUnique crowd-sourced experiment in January 2021.
On the installation page, users are informed of its purpose and of the data that is collected.
To safeguard users' privacy, collected traces are only associated with a random identifier created when the extension is installed, and participants can delete all their data by submitting their extension ID.
Out of an abundance of caution, we decided not to publish the weights of the triplet loss model trained on these users, since it can enable an attacker to track these users.
The extension and the handling of collected data conform to the IRB recommendations we received.

\subsection{Fingerprinting countermeasures}\label{sec:countermeasures}

Countermeasures can be divided into three groups.

\parhead{Blocking Scripts.} Filter lists block resources known to be a threat to user privacy.
This is the case of Brave's Shield mechanism~\cite{braveshields} and extensions, such as Ghostery~\cite{Ghostery} or Privacy Badger~\cite{PrivacyBadger}. 
However, filter lists against trackers and fingerprinting have been shown to lack exhaustiveness~\cite{FouadBLS20,DBLP:journals/corr/abs-2008-04480}.

\parhead{API Blocking.} Tor Browser, by default, and Firefox, with specific configuration, prevent web pages from reading out the contents of the canvas for privacy reasons. 
Our technique does not examine the canvas content, but rather measures the time required to draw different graphics primitives.
Snyder \etal~\cite{SnyderTK17} consider the WebGL specification a ``low-benefit, high-cost standard'', which is required by less than 1\% of the Alexa Top\,10k websites.
This may lead some people to consider the extreme option of completely blocking WebGL, as possible way of preventing GPU fingerprinting.
Disabling WebGL, however, would have a non-negligible usability cost, especially considering that many major websites rely on it, including Google Maps, Microsoft Office Online, Amazon and IKEA.
As a form of compromise, we note that Tor Browser currently runs WebGL in a ``minimum capability mode'', which allows some WebGL functionality while preventing access to the \texttt{ANGLE\_instanced\_arrays} API used by our attack. 

\parhead{Changing Attribute Values.} Defenses can change an attribute value either to make it similar with common values shared by a large proportion of users, or to add noise to it. 
For example, Tor Browser unifies the values of many attributes for all users so that their fingerprint is identical, and some browser extensions add noise to rendered canvas images~\cite{DBLP:conf/uss/VastelLRR18}.
Wu \etal~\cite{DBLP:conf/uss/WuLCW19} introduced a countermeasure that eliminates the differences in floating point operations during the rendering process to eliminate the differences in the rendering composition of WebGL.
Blurring defenses on canvas and WebGL focus on changing values.
Our technique does not directly rely on the differences in images in a rendering process, and therefore is not affected by the countermeasure of Wu \etal~\cite{DBLP:conf/uss/WuLCW19}.

There are three elements that are crucial to our fingerprinting technique: the ability to issue drawing operations in parallel.
The entire graphics stack tendency to deterministically choose which EU will render each vertex.
And the ability to measure the time it takes to render.
Disrupting any of these elements could affect the accuracy of our technique. 

\parhead{Preventing Parallel Execution.} To block our method, graphics stack could limit each web page to a single EU, or disable hardware-accelerated rendering altogether and use a deterministic software-only pipeline~\cite{DBLP:conf/uss/WuLCW19}.
However, this would severely affect usability and responsiveness, because WebGL is built around massive parallelism. 
Existing graphics APIs do not also support partitioning execution to a subset of EUs at the moment.

\parhead{Preventing Deterministic Dispatching.} Adding a randomization step to the GPU's dispatcher would make it impossible for the web page to choose which EU receives which vertex.
Assuming the dispatcher still attempts to fill up all available EUs, the effect on performance can be minimized.
We note that this countermeasure is not perfect, since a permuted trace still contains data about the system being fingerprinted.

\parhead{Preventing Time Measurements.} 
Countermeasures that reduce, or even disable, the availability of timer APIs can affect our technique, but completely blocking timing measurements from the web is known to be a futile task~\cite{DBLP:conf/fc/SchwarzMGM17,PP0}.

\subsection{Limitations and Insights}\label{s:limitations_and_observations}
\parhead{Experimental Limitations.}
The in the wild, crowd-sourced experiments demonstrate that \DrawnApart can work successfully in a variety of conditions that are not under the attacker's control.
However, our lab experiments only cover a limited set of conditions.
Specifically, 
we only evaluated the impact of temperatures between \SI{26.4}{\celsius} and \SI{37}{\celsius},
demonstrating no impact on the results.
Hence, we cannot preclude the possibility that temperatures outside this range do not affect the results.
Similarly, our lab experiments do not control for GPU voltage variations, which could affect our fingerprinting capability.
These limitations notwithstanding, the results of the crowd-sourced experiments do
provide confidence that \DrawnApart is effective in normal operating conditions.

\parhead{Approach Limitations.} 
We evaluate the effect of device restarts on fingerprinting accuracy by training a model on the \GenIII devices, and testing the model against traces collected after rebooting the devices.
We obtain an overall accuracy of 50.3\%. 
We observe that the accuracy drop is not uniform.
That is, some devices maintain stable fingerprints across restarts, whereas the fingerprints of others change significantly each restart.
We note that we do not track reboots in our in-the-wild experiments.
Hence, these already account for the potential accuracy drop associated with restarts.

We evaluate our technique across ten Chrome versions, from \textsf{80.0.3987.116} to \textsf{81.0.4044.138}.
These ten versions consist of two groups: the \textsf{v80} group which includes six minor versions, and the \textsf{v81} which includes four minor versions.
We train our classifier on the latest \textsf{v80} version (\textsf{80.0.3987.163}) and test all ten versions.
We obtain an accuracy of around 90\% on all \textsf{v80} versions, but significantly lower accuracy, of around 60\%, when we test the trained model on \textsf{v81}.
We hypothesize some changes in Chrome between \textsf{v80} and \textsf{v81} affected the entire WebGL stack. 
Observing the changelog for the Chromium code repository reveals more than 10,000 commits between the two versions with several hundreds affecting the GPU and the WebGL API~\cite{chrome8081}.
An additional experiment we conducted show that an attacker with a limited trace capture budget can maintain an up-to-date classification model by training a combined model with traces from multiple versions and obtaining a consistently high accuracy of 90$\pm$\% across all ten versions. 

\subsection{Future Work}
\parhead{In-depth Root Cause Analysis.} 
We shared our work with a committee of WebGL experts in an effort to investigate the root cause of \DrawnApart.
They acknowledged that the results reported in the paper offer insight on the tracking implications that WebGL can introduce and that our method can highlight differences introduced by the hardware manufacturing process. 
They propose additional hypotheses for the mechanism through which manufacturing variations enable \DrawnApart. Specifically, the two propsals are that:
\begin{enumerate*}
  \item DrawnApart might be uncovering differences in power consumption. 
A study by von Kistowski et al.~\cite{vonKistowski} noticed differences in power consumption from identical CPUs under the same load but it remains to be seen if and how this could translate to GPUs and WebGL.
  \item The effect might be induced by a difference in the response to temperature curves.
\end{enumerate*} 
Validating either hypothesis requires detailed knowledge of the design and the manufacturing process, which are only available to the manufacturers, and are likely beyond the scope of a typical academic research.

\parhead{Next-Generation GPU APIs.}
\DrawnApart currently only uses the WebGL API, limiting its speed and accuracy. 
Upcoming Web-based compute-specific GPU interfaces may allow for far more efficient fingerprinting. 
There are two compute-specific GPU APIs for web browsers: \WebGLtwo Compute and WebGPU.
\WebGLtwo Compute was integrated into Chrome but disabled in 2020~\cite{webgl_compute_enable}, and its development has been subsumed by WebGPU~\cite{nocompute}.
WebGPU is currently under active development, and is not supported in the stable edition of any browser,
but preliminary implementations can be found in the canary versions of Firefox, Chrome, and Edge. 

These APIs introduce \emph{compute shaders}, a form of computational pipeline that coexists with the existing graphics pipeline.
One significant feature offered to compute shaders is the ability to synchronize among different work units, by using atomic functions, message queueing or shared memory.
We used this synchronization primitive to prototype a faster fingerprinting technique for \WebGLtwo Compute.
In our prototype, all workers race to acquire a mutex, and we record the order in which the different work units were granted the mutex.
We tested this fingerprinting technique on our \GenIII corpus, after enabling \WebGLtwo Compute support in Chrome through a command-line parameter.
This compute-based fingerprint delivered a near-perfect classification accuracy of 98\%, 
while taking only 150 milliseconds to run, much faster than the onscreen fingerprint which took a median time of 8 seconds to collect.
We believe that a similar method can also be found for the WebGPU API once it becomes generally available.
The effects of accelerated compute APIs on user privacy should be considered before they are enabled globally.

\section{Related work}\label{sec:relatedwork}
\parhead{Web-based Fingerprinting.}
Eckersley~\cite{DBLP:conf/pet/Eckersley10} was the first to show that it is possible to fingerprint browsers based on their features and configurations. 
Mowery \etal~\cite{mowery2011fingerprinting} classified web fingerprinting use as constructive or destructive.
Constructive fingerprinting can detect bots~\cite{DBLP:conf/ccs/BurszteinMPT16, DBLP:conf/esorics/JonkerKV19, vastel20}, or help to protect sign-in processes~\cite{DBLP:conf/acsac/AlacaO16, DBLP:conf/dimva/LaperdrixABN19}.
Conversely, destructive use can track users and their browsing habits.
Many browser attributes are considered parts of a browser fingerprint, including \texttt{navigator} and \texttt{screen} properties~\cite{DBLP:conf/pet/Eckersley10, DBLP:conf/sp/LaperdrixRB16}, font enumeration~\cite{DBLP:conf/sp/NikiforakisKJKPV13}, audio rendering~\cite{DBLP:conf/ccs/EnglehardtN16}, and the WebGL canvas~\cite{mowery2012pixel}.
These techniques are all unable to tell apart identical devices.

\parhead{Mobile Fingerprinting.}
Mobile devices have less hardware and software diversity compared to desktops~\cite{DBLP:conf/www/Gomez-BoixLB18}.
However, they possess additional fingerprinting sources such as sensors~\cite{DBLP:conf/sp/ZhangBS19,DeyRXCN14,DasBC18,DasABP18,DBLP:journals/corr/BojinovMNB14}, microphones~\cite{clarkson12breaking,das2014poster,das14acoustic,zhou14acoustic,baldini19acoustic} and cameras~\cite{zhongjie18camera,cozzolino20noiseprint}.
Manufacturing variations can also manifest as differences in the radio frequency (RF) behavior of networked devices~\cite{AlShawabkaRDJR20, AgadakosAPA20}.
These techniques are tailored to mobile and RF environments, while our technique works in all browsers that support WebGL, without requiring permissions, additional sensors or RF hardware.

\parhead{Physically Unclonable Functions.}
The silicon-based physically unclonable function (PUF) concept is based on the idea that, even if a set of several integrated circuits is created through an identical manufacturing process, each circuit is actually slightly different due to normal manufacturing variability.
This variability can be used as a unique device fingerprint based on hardware.
Examples of silicon PUF sources include logic race conditions~\cite{DBLP:journals/pieee/HerderYKD14, DBLP:conf/dac/SuhD07}, Rowhammer behavior~\cite{DBLP:journals/cryptography/Anagnostopoulos18a}, and SRAM initialization data~\cite{DBLP:journals/tc/HolcombBF09, holcomb2007initial}.
Ruhrmair \etal~\cite{ruhrmair2012security} defined a fingerprint as ``a small, fixed set of unique analog properties'', and explain that the fingerprint should be measured quickly and preferably by an inexpensive device.
In this work the GPU is used as a PUF, and our challenge is how to successfully capture the PUF behavior while using the limited APIs available to a web browser.

\section{Conclusion}\label{sec:conclusion}
We introduced an effective technique to create a browser fingerprint that relies on minor manufacturing variations in GPUs.
To the best of our knowledge, this is the first time hardware features have been used to challenge privacy in this context.
Our fingerprinting technique can tell apart devices that are completely indistinguishable by current state-of-the-art methods, while remaining robust to changing environmental conditions.
Our technique works well both on PCs and mobile devices, has a practical offline and online runtime, and does not require access to any extra sensors such as the microphone, camera, or gyroscope. 

Processor designs are increasingly relying on massively parallel architectures to improve performance without breaking the physically-imposed constraints of power consumption and processor speed.  
As the capabilities of GPU hardware become increasingly exposed to untrusted web applications through APIs such as WebGPU, hardware and software designers must be aware of the risks to privacy raised by hardware fingerprinting, and take care to design software, drivers and hardware stacks in ways that protect user privacy.

\parhead{Responsible Disclosure.}
We shared a preliminary draft of our paper with Intel, ARM, Google, Mozilla and Brave during June-July 2020 and continued sharing our progress with them throughout 2020 and 2021.
In response to the disclosure, the Khronos group responsible for the WebGL specification has established a technical study group to discuss the disclosure with browser vendors and other stakeholders.

\parhead{Artifact Availability.} 
The JavaScript and GLSL collection code in the online, offline and GPU-based methods, the machine learning pipeline, as well as the \GenIII, \GenIV, \GenVIII and \GenX datasets, are all available in the following repository: \url{https://github.com/drawnapart/drawnapart}.
The repository includes an interactive Python notebook, viewable over the web, that demonstrates classification over real data.


\begin{acks}

This research has been supported by
  ANR-19-CE39-0007 MIAOUS;
  ANR-19-CE39-00201 FP-Locker projects;
  an ARC Discovery Early Career Researcher Award DE200101577; 
  an ARC Discovery Project number DP210102670; 
  the Blavatnik ICRC at Tel-Aviv University; 
  Intel Corporation; 
  and
  Israel Science Foundation.

  We thank Gil Fidel, Anatoly Shusterman and Antoine Vastel for their advice and help.
  We are grateful to the BGU SISE technical support engineers Vitaly Shapira and Sergey Korotchenko for their help in setting up the evaluation test-beds.
  Experiments presented in this paper were carried out using the Grid'5000 testbed, supported by a scientific interest group hosted by Inria and including CNRS, RENATER and several Universities as well as other organizations (see \url{https://www.grid5000.fr)}. 
  Parts of this work were carried out while Yuval Yarom was affiliated with CSIRO's Data61.
\end{acks}

\bibliographystyle{IEEEtranS}

\appendices
\crefalias{section}{appendix}

\section{Evaluation of \GenIII, \GenIV, and \GenVIII devices}\label{app:detailedeval}

We report the complete evaluation for 600 traces, 20 points, and 5 iterations per point in the online setting, for the \GenIII (\cref{t:results_gen3_20x5}), \GenIV (\cref{t:results_gen4_20x5}), and \GenVIII (\cref{t:results_gen8_20x5}) datasets. 

\begin{table}[h]
	\centering
	\caption{Evaluation for the \GenIII dataset, depending on the operators. The baseline is 10\%}\label{t:results_gen3_20x5}
	\begin{filecontents*}{figures/gen3_600_traces_performance_now_20_points_5_iterations.csv}
Operator,Generation,Total Dataset Size,Accuracy,Standard Deviation,Entropy,Corpus Size,Median Time
\texttt{mul},GEN3,6000,$93.5\%\pm 0.7\%$,0.008,3.22496636500027,10,3097
\texttt{sinh},GEN3,6000,$89.5\%\pm 1.4\%$,0.016,3.16242489893156,10,8757
\texttt{abs},GEN3,6000,$88.4\%\pm 1.0\%$,0.011,3.14431834530625,10,4532
\texttt{pow},GEN3,6000,$87.5\%\pm 0.3\%$,0.003,3.12928301694497,10,4663
\texttt{log},GEN3,6000,$87.1\%\pm 1.2\%$,0.014,3.12212049105055,10,9839
\texttt{exp2},GEN3,6000,$87.0\%\pm 0.6\%$,0.007,3.12018602753186,10,7532
\texttt{shl},GEN3,6000,$86.3\%\pm 1.7\%$,0.019,3.10852445677817,10,6799
\texttt{atanh},GEN3,6000,$81.8\%\pm 1.4\%$,0.016,3.03268861870617,10,11184
\texttt{inversesqrt},GEN3,6000,$80.1\%\pm 0.7\%$,0.008,3.00120174520975,10,6799
\texttt{trunc},GEN3,6000,$67.1\%\pm 1.4\%$,0.016,2.74631276642546,10,1667
\end{filecontents*}
\pgfplotstabletypeset[
	col sep=comma,
	columns={Operator,{Accuracy},{Median Time}},
	columns/Operator/.style={string type, column type={l}},
	columns/Accuracy/.style={string type,column type={l}},
	columns/Median Time/.style={column name=Median time (ms), column type={R{1.2cm}}},
	every head row/.style={before row=\toprule,after row=\midrule},
	every last row/.style={after row=\bottomrule}
	]{figures/gen3_600_traces_performance_now_20_points_5_iterations.csv}
\end{table}

\begin{table}[h]
	\centering
	\caption{Evaluation for the \GenIV dataset, depending on the operators. The baseline is 4\%}\label{t:results_gen4_20x5}
	\begin{filecontents*}{figures/gen4_600_traces_performance_now_20_points_5_iterations.csv}
Operator,Generation,Total Dataset Size,Accuracy,Standard Deviation,Entropy,Corpus Size,Median Time
\texttt{mul},GEN4,13800,$32.7\%\pm 0.3\%$,0.004,2.91137201269596,23,6361
\texttt{abs},GEN4,13800,$29.3\%\pm 0.5\%$,0.006,2.75060650483559,23,3295
\texttt{shl},GEN4,13800,$28.7\%\pm 0.4\%$,0.004,2.72246602447109,23,6483
\texttt{inversesqrt},GEN4,13800,$28.2\%\pm 0.7\%$,0.008,2.69488019279919,23,6485
\texttt{exp2},GEN4,13800,$27.8\%\pm 1.0\%$,0.012,2.6773203056995,23,6528
\texttt{trunc},GEN4,13800,$26.6\%\pm 1.1\%$,0.013,2.61549477070742,23,3161
\texttt{log},GEN4,13800,$25.3\%\pm 1.0\%$,0.011,2.54267077900472,23,7673
\texttt{pow},GEN4,13800,$23.3\%\pm 0.6\%$,0.006,2.42447425488693,23,9370
\texttt{sinh},GEN4,13800,$19.5\%\pm 0.5\%$,0.005,2.16242489893156,23,8953
\texttt{atanh},GEN4,13800,$19.1\%\pm 0.6\%$,0.006,2.13641015875566,23,10099
\end{filecontents*}
\pgfplotstabletypeset[
	col sep=comma,
	columns={Operator,{Accuracy},{Median Time}},
	columns/Operator/.style={string type, column type={l}},
	columns/Accuracy/.style={string type,column type={l}},
	columns/Median Time/.style={column name=Median time (ms), column type={R{1.2cm}}},
	every head row/.style={before row=\toprule,after row=\midrule},
	every last row/.style={after row=\bottomrule}
	]{figures/gen4_600_traces_performance_now_20_points_5_iterations.csv}
\end{table}

\newpage
\pagebreak

\begin{table}[h]
	\centering
	\caption{Evaluation for the \GenVIII dataset, depending on the operators. The baseline is 6\%}\label{t:results_gen8_20x5}
	\begin{filecontents*}{figures/gen8_600_traces_performance_now_20_points_5_iterations.csv}
Operator,Generation,Total Dataset Size,Accuracy,Standard Deviation,Entropy,Corpus Size,Median Time
\texttt{exp2},GEN8,10200,$43.6\%\pm 1.2\%$,0.013,2.88979799885563,17,3172
\texttt{inversesqrt},GEN8,10200,$39.4\%\pm 0.9\%$,0.010,2.74487867580947,17,3181
\texttt{pow},GEN8,10200,$36.6\%\pm 1.4\%$,0.016,2.63768752173032,17,4698
\texttt{log},GEN8,10200,$33.6\%\pm 0.5\%$,0.005,2.51559472006177,17,3299
\texttt{sinh},GEN8,10200,$32.4\%\pm 0.9\%$,0.011,2.46248863969523,17,4569
\texttt{abs},GEN8,10200,$30.9\%\pm 0.6\%$,0.007,2.3932331289605,17,3174
\texttt{mul},GEN8,10200,$30.9\%\pm 1.0\%$,0.011,2.39094277280254,17,3173
\texttt{atanh},GEN8,10200,$30.6\%\pm 0.5\%$,0.005,2.3775865222019,17,5935
\texttt{trunc},GEN8,10200,$28.7\%\pm 0.6\%$,0.007,2.28885069763793,17,3174
\texttt{shl},GEN8,10200,$26.9\%\pm 1.1\%$,0.013,2.19482250457998,17,3172
\end{filecontents*}
\pgfplotstabletypeset[
	col sep=comma,
	columns={Operator,{Accuracy},{Median Time}},
	columns/Operator/.style={string type, column type={l}},
	columns/Accuracy/.style={string type,column type={l}},
	columns/Median Time/.style={column name=Median time (ms), column type={R{1.2cm}}},
	every head row/.style={before row=\toprule,after row=\midrule},
	every last row/.style={after row=\bottomrule}
	]{figures/gen8_600_traces_performance_now_20_points_5_iterations.csv}
\end{table}

\section{Deterministic attributes collected for the in-the-wild dataset}\label{app:attributes}

\lstset{ 
  basicstyle=\footnotesize\ttfamily,        
  breakatwhitespace=false,         
  breaklines=true,                 
  language=Java,                 
  showspaces=false,                
  showstringspaces=false,          
  showtabs=false,                  
}
\begin{lstlisting}
cookies and session support,
HTTP headers: [Accept, Accept-Encoding, Language, User-Agent],
navigator: [DNT, platform, plugins],
screen: [width, height]
timezone,
WebGL: [vendor, renderer]
\end{lstlisting}

\section{Selected Hyperparameters}\label{a:hyperparameters}
\cref{t:hyperparameters} summarizes the hyperparameters for the classifiers used in this work.
\begin{table}[htb]
\centering
  {\footnotesize 
  \caption{Hyperparameters for the CNN classifier\label{t:hyperparameters}}
\begin{tabular}{lll}
\toprule
  \textbf{Hyperparameter} & \textbf{Value} & \textbf{Space} \\
\midrule
Embedding size                  & 256   & 32--256 \\
Number of convolution blocks    & 3  &  1--10 \\
Batch size                      & 32    & 32--1024      \\
Convolution filter size         & 128    & 8--128      \\
Convolution kernel size         & 4    & 2--5      \\
Dropout rate                    & 0.119510    & 0--0.5      \\
Activation                      & relu   & relu, sigmoid   \\
\bottomrule
\end{tabular}
  }
\end{table}

\section{Evaluation of the standalone pipeline on additional browsers in the wild}\label{app:addl-browsers-in-the-wild}

\begin{table}[htb]
	\centering
  	\caption{Standalone Performance of \DrawnApart over multiple browsers}
  	\label{t:wild_eval_browsers}
	\begin{tabular}{llrrr}
		\toprule
  		\multirow{2}{*}{\textbf{Browser}} & \multicolumn{3}{c}{Accuracy (\textit{Base rate})} \\
	  & \multicolumn{1}{c}{\textbf{Top-1}} 	& \multicolumn{1}{c}{\textbf{Top-5}} 	& \multicolumn{1}{c}{\textbf{Top-10}}\\
		\midrule
		Chrome                  &  24.31\% (\textit{0.7\%}) 			& 49.12\% (\textit{2.9\%}) 			& 60.80\% (\textit{4.7\%}) \\
		Edge                    &  52.60\% (\textit{2.9\%})   		& 85.48\% (\textit{15.6\%})			& 93.86\% (\textit{29.7\%}) \\
    Opera                   &  79.28\% (\textit{17.9\%})    		& 99.41\% (\textit{50.7\%})			& 100.0\% (\textit{77.5\%})\\
    Yandex                  &  89.69\% (\textit{27.6\%})   		& 98.36\% (\textit{85.9\%}) 			& 99.76\% (\textit{94.1\%})\\
		\bottomrule
	\end{tabular}
\end{table}

\newpage
\pagebreak

\section{\FPStalker Hybrid algorithm with Drawn Apart addition}\label{app:hybridfpwithda}

\begin{algorithm}[h]
    \label{alg:hybridfpwithda}
    \caption{Hybrid matching algorithm with the \DrawnApart addition highlighted in red}

    \SetAlgoLined
    \SetKwProg{Fn}{Function}{}{}
    \Fn{FingerprintMatching (F, $f_{u}$, $\lambda$, $\epsilon$)}{
      \For(){$f_{k}$ $\in$ F}{
        \lIf(){FingerPrintHasDifferences($f_{k}$, $f_{u}$, rules)}{
          $F_{ksub}$ $\leftarrow$ exact $\cup$ $<f_{k}>$
        }
        \Else(){
          exact $\leftarrow$ exact $\cup$ <$f_{k}$>
        }
      }
      \If(){$\left | exact \right |$ $>$ 0}{
        \lIf(){SameIds(exact)}{
          \Return{exact[0].id}
        }
        \lElse(){
          \Return{GenerateNewId()}
        }
      }
  
      \For(){$f_{k}$ $\in$ $F_{ksub}$}{
        \tikzmk{A}cosine\_sim $\leftarrow$  GetSimilarity($f_{u}.avg\_embedding$, $f_{k}.avg\_embedding$)\;
        \If(){cosine\_sim $>$ $\epsilon$}{
        	\Return{$f_{k}.id$}
        }
        \tikzmk{B}
		\boxit{red}
        $<x_{1}, x_{2}, .. , x_{m}>$ = FeatureVector($f_{u}$, $f_{k}$)\;
        p $\leftarrow$ P($f_{u}.id$ = $f_{k}.id$ $|$ $<x_{1}, x_{2}, .. , x_{m}>$)
  
        \If(){p $\geq$ $\lambda$ }{
            candidates $\leftarrow$ candidates $\cup$ $<f_{k}, p>$
        }
      }
  
      \If(){$\left | candidates \right | >$ 0}{
        \lIf(){$|$GetRankAndFilter(candidates)$| >$ 0}{
          \Return{candidates[0].id}
        }
      }
      \Return{GenerateNewId()}
    }
\end{algorithm}

\end{document}